\documentclass[preprint2]{emulateapj}
\usepackage{natbib}
\newcommand{\hii}{\mbox{H{\sc ii}~}}

\begin{document}

\title{STAR FORMATION IN W3 - AFGL333: YOUNG STELLAR CONTENT, PROPERTIES AND ROLES OF EXTERNAL FEEDBACK}

\author{Jessy Jose  \altaffilmark{1}, Jinyoung S. Kim\altaffilmark{2}, Gregory J. Herczeg\altaffilmark{1}, Manash R. Samal\altaffilmark{3}, John H. Bieging\altaffilmark{2},  Michael R. Meyer\altaffilmark{4} and William H. Sherry\altaffilmark{5,6}}
\altaffiltext{1}{Kavli Institute for Astronomy and Astrophysics, Peking University, Yi He Yuan Lu 5, Haidian Qu, Beijing 100871, China; jessyvjose1@gmail.com}
\altaffiltext{2}{Steward Observatory, University of Arizona, 933 North Cherry Avenue, Tucson, AZ 85721-0065, USA}
\altaffiltext{3}{Aix Marseille Universit\'{e}, CNRS, LAM (Laboratoire d'Astrophysique de Marseille) UMR 7326, 13388 Marseille, France}
\altaffiltext{4}{Institute for Astronomy, ETH Zurich, Wolfgang-Pauli-Strasse 27, 8093 Zurich, Switzerland}
\altaffiltext{5}{National Optical Astronomy Observatories, 950 North Cherry Avenue, Tucson, AZ 87719, USA}
\altaffiltext{6}{Eureka Scientific, Inc. 2452 Delmer Street Suite 100 Oakland, CA 94602-3017}

\begin{abstract}

One of the key questions in the field of star formation is the  role of stellar feedback 
on subsequent star formation process. The W3 giant molecular cloud complex at the
western border of the W4 super bubble  is thought to be  influenced by the stellar winds of the massive stars in W4.
AFGL333 is a $\sim$ 10$^4$ $M_{\odot}$ cloud within  W3.
This paper presents a study of the star formation activity within AFGL333 using deep $JHK_s$ photometry 
obtained from the NOAO Extremely Wide-Field Infrared Imager  
combined with {\it Spitzer}-IRAC-MIPS photometry. Based on the infrared excess, we identify 812 candidate young stellar objects 
in the complex,  of which 99 are classified as Class I and 713 are classified as Class II sources. 
The stellar density analysis of  young stellar objects  reveals  three major stellar aggregates within AFGL333, 
named here AFGL333-main, AFGL333-NW1 and AFGL333-NW2.  The disk fraction  within AFGL333 is estimated
to be   $\sim$ 50--60\%. 
We use   the extinction map made from  the $H-K_s$ colors of the background stars to understand the cloud structure and to
estimate the cloud mass.  The CO-derived extinction map corroborates the cloud structure and mass estimates from NIR color method. 
From the stellar mass and cloud mass associated with AFGL333, we infer that
the region is currently forming stars with an efficiency of  $\sim$ 4.5\% and  at a rate of $\sim$
2 - 3 $M_{\odot}$ Myr$^{-1}$pc$^{-2}$. In general, the star formation activity within AFGL333 is 
comparable to that of nearby low mass star-forming regions.  We do not find any strong evidence to suggest
that the stellar feedback from the massive stars of nearby W4 super bubble has affected the global star
formation properties of  the  AFGL333 region. 

\end{abstract}


\section{Introduction}
\label{intro}

Most stars form in OB associations, where  stellar winds, UV radiation and  supernova
explosions from the massive members significantly affect their environment.  Feedback from newly
born massive stars   plays  an active  role in the subsequent evolution of their parent molecular
cloud.  Feedback can trigger the birth of new generations of stars that would otherwise not exist,
allowing star formation to propagate continuously from one point to the next \citep{elmegreen1977}.
On the other hand, feedback may also inhibit star formation by clearing away the dust and gas
\citep{bisbas2011,walch2013,dale2013}. 

Numerical simulation studies by \citet{dale2012} and \citet{dale2013} show that the effect of feedback  in a massive
star forming region depends on its   cloud mass and density. The most massive and largest clouds are
mostly dynamically unaffected by stellar feedback. On the other hand, feedback has a profound effect
on the lower density clouds, expelling tens of percent of the neutral gas long before any  massive star
explode as a supernova. Since giant molecular clouds typically have  complicated, clumpy internal structures, 
the stellar feedback mechanisms penetrate into different depths in different directions of a given cloud,
producing highly irregular morphologies \citep{walch2013}. The  global influence of feedback on a given
system may  therefore differ  from the  local effects; star formation could be suppressed at some locations and
triggered in another location \citep{dale2012}.  

\begin{figure}[h]
\centering 
 \includegraphics[scale = 0.55, trim =15  15 20 40,clip]{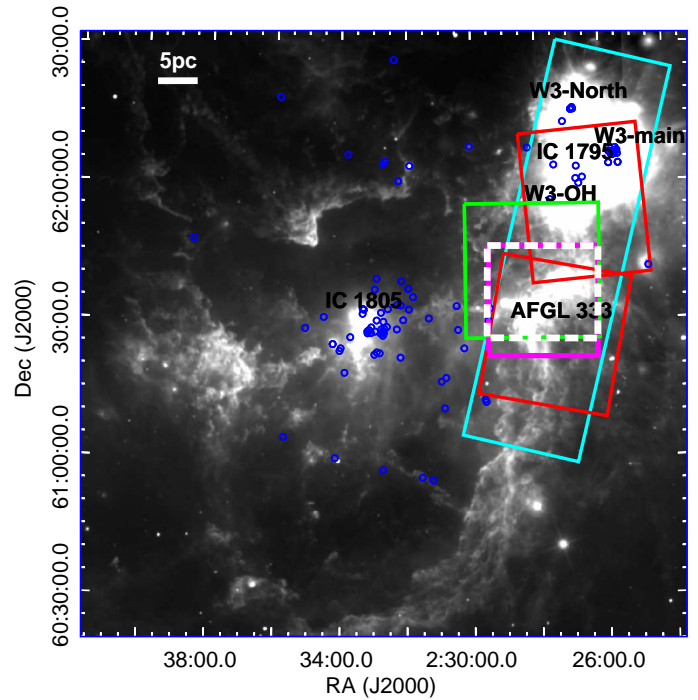}
\caption{WISE 22 $\mu$m wide field image ($\sim$ 2$^\circ$$\times$2$^\circ$)  
covering W4 super bubble centred at the cluster IC 1805, along with  W3 GMC at its west. 
Locations of the important sub-regions of the W3 complex are marked and the boxes represent the area covered for various observations 
(green: NEWFIRM; red: IRAC; cyan: MIPS). Magenta box shows the area ($\sim$ 24$^\prime$ $\times$ 24$^\prime$)  corresponding to AFGL333 region, 
based on the low  resolution $^{12}$CO(J=3-2) map by  \citet{sakai2006}  and the white dashed box represents the area considered
in this study.  The blue circles denote the locations of the
probable massive members (earlier than B3V) in the complex obtained from SIMBAD database\footnote{http://simbad.u-strasbg.fr/simbad/}.  
 }
 
\label{w4}
\end{figure}

The W3 star forming complex  (d $\sim$ 2 kpc; \citealt{xu2006,hachisuka2006}), one of the most
massive molecular clouds in the outer Galaxy  ($\sim$ 4 $\times$ 10$^5$ $M_{\odot}$; \citealt{moore2007,polychroni2012}),
has long been proposed as a classic example of induced or triggered  star
formation (\citealt{lada1978,oey2005}).  The W3 complex has a complicated  structure. The pressure from the expanding \hii region  and
stellar winds from the massive stars of W4 super bubble have been suggested to  have swept up the molecular cloud to 
create the `high density layer' (HDL) at its western periphery (\citealt{lada1978} and references therein). W3 Main, W3 (OH), 
W3 North, IC1795  and AFGL333  are the most active star forming sites identified within the high density layer. 
Feedback from W4  was identified as a key factor for inducing and enhancing the star formation activity within the high density layer.
Localized triggering from IC1795  has been suggested to influence its surrounding regions such as W3 Main and W3 (OH).  Whereas, to the west of HDL
(e.g., KR140), spontaneous or quiescent mode of star formation has been suggested often
(e.g.,  \citealt{kerton2008,ingraham2011,ingraham2013}).   
The most prominent regions within W3 are marked  in Fig. \ref{w4}. Of these, W3 Main is the most active star forming region, with more 
than 10  \hii regions of various evolutionary status (\citealt{tieftrunk1997,ojha2004a}). 
The scenario called  `convergent constructive feedback'
proposed by \citet{ingraham2013,ingraham2015} suggests that   star formation activity towards W3 Main is a self-enhancing process,
where the youngest and most massive stars are observed at the innermost regions.

\begin{figure*}[t]
\centering

 \includegraphics[scale = 0.7, trim =10  0 0 10,clip]{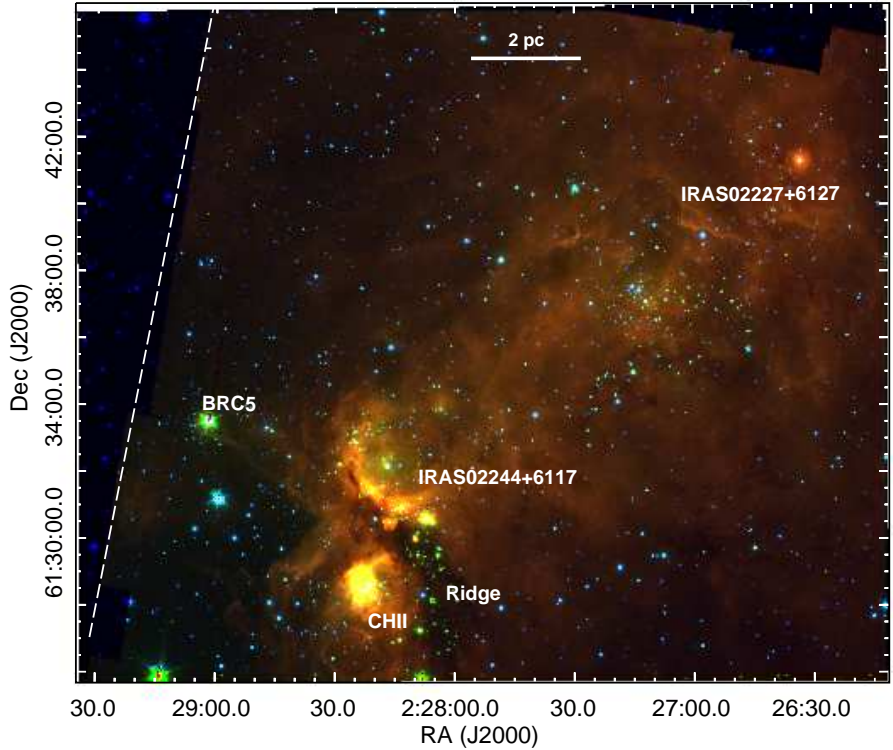}
\caption{ Color composite image of  $\sim$ 24$^\prime$$\times$20$^\prime$ area considered in this study
(centered at $\alpha_{2000}$ = $02^{h}27^{m}48^{s}$; $\delta_{2000}$ = $+61^{\circ}35^{\prime}58^{\prime\prime}$) made using
 2.1  $\mu$m (blue);   4.5 $\mu$m (green) and  8.0 $\mu$m (red) images. The important sub-regions are marked and 
the area to the left of dashed line is not covered by 4.5 and 24 $\mu$m bands.
}

\label{area}
\end{figure*}

Within the W3 complex,  the AFGL333 region ($\alpha_{2000}$ = $02^{h}28^{m}15^{s}$; $\delta_{2000}$ =
$+61^{\circ}20^{\prime}58^{\prime\prime}$)  lies at the  southern part of the HDL, $\sim$ 17 pc away from the centre of
W4 bubble  (see Fig. \ref{w4}). AFGL333 consists of 1) a bright rimmed cloud  (BRC 5, \citealt{sugitani1991}) associated with  IRAS 02252+6120,
pointing directly to  the massive stars of W4, 2) an \hii region  ionized by a B0.5 star \citep{hughes1982}  associated with IRAS 02245+6115,
3) a prominent dense filamentary structure associated with a molecular ridge (defined as AFGL333 Ridge; \citealt{ingraham2013})
and several IRAS sources (see Fig. \ref{area}).  
The morphology of this complex with BRC 5  facing the W4 OB association,  strongly suggests a large
scale feedback due to the expansion of W4. The projected distance between  AFGL333 and the massive stars in W4 is smaller when
compared to that of the distance between W3 Main and W4. Hence the amount of stellar radiation and wind energy received 
at the surface of AFGL333-main may be  higher than that of  W3 Main.  

 The  high density layer of the  W3 complex has diverse density structure, where W3 Main has the highest density compared to AFGL333 \citep{ingraham2015}. 
Is the outcome of star formation process such as star formation efficiency, star formation rate etc.  different in AFGL333 and  
W3 Main?  Stars form in both regions, but from gas with different densities subjected to different  radiation environments.
 AFGL333 seems to be externally influenced by W4, whereas the local triggering effect from the cluster IC1795 and the internal self-enhancing 
process have been given as the explanation for the high star formation activity within W3 Main (\citealt{oey2005,ingraham2013}). 
In order to understand the nature of YSO population and star formation activities
in the  sub-regions of W3, under different external conditions,  we explore the stellar properties  of this complex.  
 The stellar content of AFGL333 - both low and high mass - is less   explored than W3 Main/OH.
 The latest census of  the young stellar  population of the W3 complex  by \citet{ingraham2011} used
 shallow 2MASS and {\it Spitzer} observations. Since the IMF of a star forming region peaks towards the low mass stars,
the star formation process is best traced by the identification of low mass stellar population. Deep near-IR  (NIR) 
and mid-IR  (MIR) photometry are ideal tools to uncover the low mass stellar content of heavily obscured and dense environments.

In this paper, we use the deep NIR data in combination with {\it Spitzer} data  to  identify and characterise the 
young stellar objects (YSOs) within AFGL333 as well as to understand the star formation activity of the region.
 The datasets presented in this study are unique in terms of its depth and completeness  compared to other surveys of this region.
The paper is organised as follows:  Section  2  discusses  the
various data  sets and  photometry catalogs  obtained. Sections 3 and 4 present detailed analysis of the extinction map and
selection procedures of YSO candidates in AFGL333. Various characteristics of YSOs and the  newly identified stellar 
aggregates within AFGL333 and their properties are discussed in Section 5. In Section 6 we
compare our results with W3 Main as well as with various nearby low and high mass  star forming regions. We also 
discuss the implications of triggered star formation in AFGL333. The results are summarised in Section 7. 

%%%%%%%%%%%%%%%%%%%%%%%%%%%%%%%%%%%%%%%%%%%%%%%%%%%%%%%%%%%%%%%%%%%%%%%%%%

%%%%%%%%%%%%%%%%%%%%%%%%%%%%%%%%%%%%%%%%%%%%%%%%%%%%%%%%%%%%%%%%%%% %%%%%

%%%%%%%%%%%%%%%%%%%%%%%%%%%%%%%%%%%%%%%%%%%%%%%%%%%%%%%%%%%%%%%%%%%%%%%%%%
%%%%%%%%%%%%%%%%%%%%%%%%%%%%%%%%%%%%%%%%%%%%%%%%%%%%%%%%%%%%%%%%%%% %%%%%

\section{Observations and point source catalogs}

\subsection{NEWFIRM-NIR imaging and photometry}
\label{newfirm}

We obtained NIR observations of AFGL333   in $JHK_s$ bands using the National Optical Astronomical Observatory (NOAO) Extremely
Wide Field InfraRed Imager (NEWFIRM; \citealt{probst2004}) camera on the  4m Mayal  telescope at  Kitt Peak National Observatory (KPNO)
on 06 December 2009. The NEWFIRM camera  contains 4 InSb
2048$\times$2048 pixel arrays arranged in a 2$\times$2 pattern with a field of view of  28$^\prime$$\times$28$^\prime$ (16.2 $\times$ 16.2 pc$^2$ at d= 2kpc) and 
a pixel scale of 0.4$^{\prime\prime}$ (0.0039 pc) per pixel. The area coverage of this observation is shown by a green box in Fig. \ref{w4}.  
We took 50, 48 and 72 second exposures for $J, H$ and $K_s$ bands, respectively at 9 dithered positions. 
 The dither offsets were  large enough to fill in the 1$^\prime$ gap between CCDs. During our observations, 
the seeing was around $\sim$ 1$^{\prime\prime}$.0 -  1$^{\prime\prime}$.2. Standard processing of dark frame subtraction, flat fielding, 
sky-subtraction and bad pixel masking was performed by the NEWFIRM Science Pipeline \citep{swaters2009}
and final stacked images were produced in each band.

 We used the DAOFIND task in IRAF to extract the preliminary list of point sources in the $K$-band image. This algorithm provides additional
statistics of the point sources such as roundness, sharpness etc. In order to avoid the artifacts and false detections, 
we selected only those sources having S/N $>$ 5. Following \citet{stetson1987}, we set the roundness 
limits as -1 to +1 and sharpness limits as  0.2 to +1, which should eliminate the bad pixel brightness enhancements and the extended 
sources such as background galaxies from the point source catalogue. Due to the non-uniform background nebular emission 
as well as crowdness of the field, some  sources were missed in the automatic detection algorithm. Such sources were manually identified
and  added in the final  source list if they satisfy  the above S/N, roundness and sharpness criteria. The list was again
visually checked for any spurious source detection  and were deleted from the list. The astrometry has been corrected with respect to the 2MASS 
point source catalog and the astrometry accuracy of our final source list is  $\sim$  0$^{\prime\prime}$.3  

The final source list was fed into the  DAOPHOT's ALLSTAR routine to obtain the  point spread function (PSF)  photometry in $J$, $H$ and $K_s$ bands.   
Absolute photometric calibration  was obtained using the Two Micron All Sky Survey (2MASS) data \citep{skrutskie2006}, with quality 
flag better than 'A' in all the three bands.   
We matched up bright isolated stars from our NEWFIRM catalogs with sources from 2MASS within a radius of 1$^{\prime\prime}$.0 and obtained
 zero points. The  rms scatter between the calibrated NEWFIRM and 2MASS data (i.e., 2MASS-NEWFIRM data) for the $J$, $H$, and $K_s$ 
bands were 0.06, 0.06 and 0.07 mag, respectively.  The saturated point sources in the NEWFIRM catalog  were 
replaced by 2MASS magnitudes.  We finally created the $JHK_s$ photometry catalog by spatially matching the detected 
sources in three bands within a match radius of 1$^{\prime\prime}$.0.  Only those sources satisfy our reliability criteria in all three bands (i.e., 
photometry uncertainty $<$0.2 mag) were included in the final list. The total number of sources detected in
each band and the detection limits  are listed in Table \ref{catalog}. Compared to the 2MASS completeness limits in $JHK_S$ bands 
(i.e.,  15.9, 15.1 and 14.3 mag), the new photometry is deeper by 4.6, 3.6 and 3.7 mag in $J$, $H$ and $K_s$ bands, respectively and 
thus adding 10444, 10624 and 11441 more sources than the existing 2MASS catalog.  

\subsection{{\it Spitzer}-IRAC-MIPS imaging and photometry}
\label{spitzer}

 Images from the {\it Spitzer} space telescope \citep{werner2004}  using Infrared Array Camera (IRAC, \citealt{fazio2004})
centered at 3.6, 4.5, 5.8 and 8.0 $\mu$m (ch1, ch2, ch3 and ch4) were obtained from the {\it Spitzer} archive\footnote 
{http://archive.spitzer.caltech.edu/}. These observations were taken  on 10 January 2004 and 19 February 2007
(Program IDs: 30955, 127; PIs: R. Gehrz, T. Moore) in high dynamic range (HDR) mode with three dithers per map 
position and two images each with integration time of 0.4 s and 10.4 s per dither. Preliminary analyses of 
these data sets have been presented in  \citet{ruch2007}  and \citet{ingraham2011}.   The total area coverage by these 
two observations is $\sim$ 28$^\prime$ $\times$ 63$^\prime$ ($\sim$ 16 $\times$  37 pc$^2$; see Fig. \ref{w4}).

We obtained the cBCD ({\it corrected basic calibrated data}) 
images (version S18.7.0)  from the  archive and the raw data was processed and calibrated with IRAC pipeline. The final mosaic images
were created using the MOPEX pipeline (version 18.0.1) with an image scale of 1$^{\prime\prime}$.2 per pixel.  In order to avoid 
the saturation due to bright nebular emission, we processed the long and short exposure frames seperately. We kept the settings  similar
to the NEWFIRM data (see Section \ref{newfirm}) to make a preliminary source list using DAOFIND in IRAF. Since the IRAC bands  suffered from variable
background nebular emission over small spatial scales,  single point source detection threshold across the entire mosaic image does not
detect all the potential point soucres in the field. We used various detection threshold values over  multiple iteration to enable 
the detection of all faint sources in the field. The reliability of the sources were decided based  on their S/N, roundness and 
sharpness values. Many spurious sources  in the nebulosity were identified by visual inspection  and deleted  
from the automated detection list. 

To extract the flux of the point sources,  we performed  point response function  (PRF) fitting on IRAC images in multi frame mode,
using the tool   Astronomical Point Source EXtraction  (APEX), developed by the {\it Spitzer}  Science Centre (see \citealt{jose2013} for details). Flux
densities were converted in to magnitudes using the  zero-points 280.9, 179.7, 115.0 and  64.1  Jys in the 3.6, 4.5, 5.8 and  8.0
$\mu$m bands, respectively, following the IRAC Data Handbook\footnote{http://ssc.spitzer.caltech.edu/irac/iracinstrumenthandbook/}. 
The saturated bright sources in the long integrated images were replaced by the sources from  short exposure images.  
 To ensure optimal photometry, only those sources with S/N $>$ 5 and photometry uncertainty $<$0.2 mag in individual bands 
are considered for further analysis.  The number of sources detected in each band and the detection limits  are given in Table \ref{catalog}.  
The IRAC data of the four band passes were merged by matching  the coordinates using a radial matching tolerance of 1$^{\prime\prime}$.2.
Thus our final IRAC catalog contains photometry of 29224  sources that are  detected in one or more IRAC bands, including  
1567  sources  detected in all four IRAC bands. 

AFGL333   was observed in 24 $\mu$m  using the  Multi band Imaging Photometer for {\it Spitzer} (MIPS; \citealt{rieke2004})  on 03 
February 2004 (Program ID: 127, PI: R. Gehrz).   The observations covers an area $\sim$ 26$^\prime$ $\times$ 89$^\prime$ ($\sim$ 15 $\times$ 52 pc$^2$,
see Fig. \ref{w4}).  We obtained the BCD images (S18.13.0) from the {\it Spitzer} archive and the final
mosaics were created using the MOPEX  pipeline (version 18.0.1) with an image scale of 2$^{\prime\prime}$.45 per pixel.
 We applied the same source detection technique and reliability criteria as described for IRAC to the MIPS 24 $\mu$m image. 
To extract the flux, we performed the PRF fitting method in the single frame mode. The zero-point value of 7.14 Jy from the 
MIPS Data Handbook\footnote{http://ssc.spitzer.caltech.edu/mips/mipsinstrumenthandbook/} has been used to convert flux densities 
to magnitudes. The final catalogue contains the 24 $\mu$m photometry of 327 sources having uncertainty $<$ 0.2 mag.

\subsection{Completeness limits}
\label{completeness}

%%%%%%%%%%%%%%%%%%%%%%%%%%%%%%%%%%%%%%%%%%%%%%%%%%%%%%%%%%%%%%%%%%%%%%%%%%%%%%%%%%%%%%%%%%%%%%%%%%
\begin{figure}[t]
\centering
\includegraphics[scale = 0.40, trim = 0 0 0 0, clip]{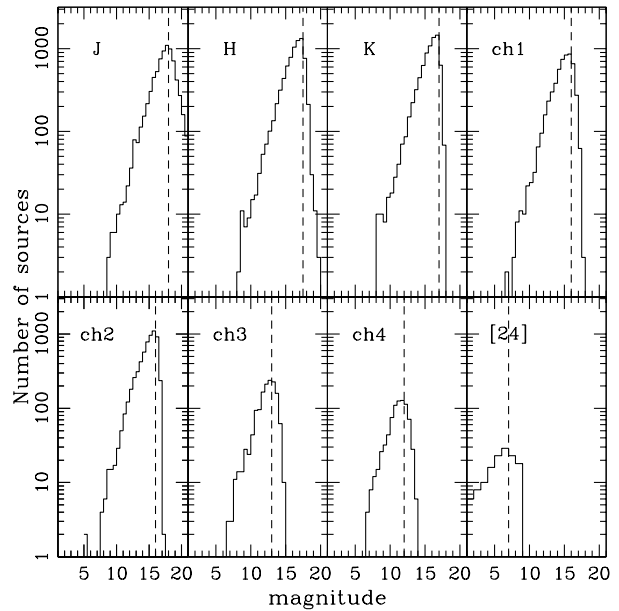}
\caption{Source histograms for bands $J$ through 24 $\mu$m showing the limiting magnitude and
completeness limit for each band. The vertical lines indicate the adopted completeness limit.  }
\label{cf}
\end{figure}
%%%%%%%%%%%%%%%%%%%%%%%%%%%%%%%%%%%%%%%%%%%%%%%%%%%%%%%%%%%%%%%%%%%%%%%%%%%%%%%%%%%%%%%%%%%%

In order to analyze the photometric incompleteness, the artificial star experiment has  been performed using  the ADDSTAR
routine  in IRAF on the $K_s$-band image (see \citealt{jose2012} for details). Briefly,   artificial stars were added at randomly generated positions  
to the $K_s$-band image. The luminosity distribution of artificial stars was chosen in such a way that more number of stars 
were inserted towards the fainter magnitude bins. The frames were reduced using the same procedure used for the original frames (Section \ref{newfirm}). 
The ratio of the number of stars recovered to those added in each magnitude interval gives the completeness factor  as a function of magnitude.
The faintest magnitude bin, where the fraction of sources recovered was greater than 90\% was adopted as the completeness limit. 
We thus obtained  $\sim$ 90\% completeness of the data  for the
magnitude limit of 17  in $K_s$-band.  The completeness limits of individual bands were also analyzed  using the   histogram 
distributions  of the measured source magnitudes (see Fig. \ref{cf}; Table \ref{catalog}). The turnover point in source count curves 
can serve as a  proxy to show the completeness limit (e.g., \citealt{jose2013,willis2013,samal2015}). 
The photometric completeness obtained from histogram analysis is consistent with that estimated from
the artificial star experiment in $K_s$-band.

%%%%%%%%%%%%%%%%%%%%%%%%%%%%%%%%%%%%%%%%%%%%%%%%%%%%%%%%%%%%%%%%%%%%%%%%%%
\begin{table}
\begin{center}
\caption{Point source catalog summary for various bands}
\label{catalog}
\centering
\begin{tabular}{ccccc}
\tableline\tableline
 Band & Sources in   & within & detection & 90\% completeness \\
 & total area$^a$ & AFGL333$^b$  & limit (mag) & limit (mag) \\

  \tableline

$J$ & 13162 & 7194 & 20.5 & 18.0 \\
$H$ & 13472 & 7447 & 18.7 & 17.5 \\
$K_s$ & 13573 & 7525 & 18.0 & 17.0\\
ch1 &  22361 & 5346 & 17.8 & 16.0\\
ch2 &  20502 & 6095 & 17.5 &16.0 \\
ch3 &  4452 & 1408 &15.3 & 13.0 \\
ch4 &  2400 & 799 & 14.0 & 12.0 \\
$[24]$& 327 &134 & 9.3 & 7.0 \\

\tableline
%%\hline
\end{tabular}
\end{center}
$^a$ Number of sources  within the field of view of individual bands.\\
$^b$ Number of sources within AFGL333 area considered in this study.\\
\end{table}
%%%%%%%%%%%%%%%%%%%%%%%%%%%%%%%%%%%%%%%%%%%%%%%%%%%%%%%%%%%%%%%%%%%%%%%%%%

\subsection{Final point source catalog}
\label{areaofinterest}

We created the final photometry  catalog by spatially matching and merging the detected sources in various bands. 
A search for the IRAC counterparts of  the NIR sources within a  match radius of $1^{\prime\prime}$.2 identified   7168  sources
at least in one of the IRAC bands.
Of the 327 detection in MIPS band, within a match radius of 2$^{\prime\prime}$.5,  304 sources  have IRAC
counterparts at least in one band and 135 sources with NIR counterparts.

We compared our IRAC and MIPS photometry with the existing catalogs in the literature (\citealt{ruch2007,ingraham2011}).
The root mean square (rms) scatter between the photometry from this work and that  of \citet{ingraham2011}
are found to be within 0.1 mag  and are  within 0.2 mag with  \citet{ruch2007} for all IRAC bands. This dispersion
occurs mainly due to the different techniques adopted for the flux estimation, however is within the  uncertainty limit  in
this study.  The rms scatter between the MIPS-24$\mu$m photometry of this study with that of \citet{ruch2007} and \citet{ingraham2011}
 is $\sim$ 0.5 mag.

Table~\ref{catalog} lists  the number of sources detected in each band as well as their detection and completeness limits. 
The number of sources in each band is different due to respective  sensitivity limits and area coverage (see Fig. \ref{w4}).
The AFGL333  region is projected on the sky over an area $\sim$ 24$^\prime$ $\times$ 24$^\prime$ (see Fig. \ref{w4}; $\sim$ 193 pc$^2$),
based on CO emission maps from \citet{sakai2006} and \citet{bieging2011}.
However, the coverage area of the near-IR, IRAC, and MIPS observations are all different and do not cover the entire region
(see Fig. \ref{w4}).  An area ( $\sim$ 24$^\prime$$\times$20$^\prime$, 161 pc$^2$)  covered by most of the wavelengths 
is considered for this analysis (see Fig. \ref{area} and white box in Fig. \ref{w4}).   Our coverage area for extinction maps (Section \ref{akmap}) and  
our YSO identification (Section \ref{yso}) miss $\sim$ 15\% of AFGL333 area to its south compared to the area delineated by \citet{sakai2006}. 
The number of sources detected within the area of interest in each band is given in column 3 of Table 1.

\section{Mapping the dust and gas in AFGL333}
\label{akmap}

 The main goal of this study is to obtain the census of young stellar objects as a tool to understand the cloud structure and
star formation activity within AFGL333 region.   The census of young stellar objects based on color-color cuts requires a 
quantification of how color-color selections are affected by extinction.   In this section, we map the structure of gas and 
dust in the AFGL333 molecular cloud. These extinction maps also allow us to identify the sites of active star formation.

The primary dust map used in this paper is a map of extinction calculated from near-IR colors of background stars (Section \ref{nirakmap}).  
In Section \ref{coakmap}, we create a map of molecular gas from 1.3 mm CO emission.  In Section \ref{akcomparison}, the near-IR extinction map is then compared 
to the CO map and a published dust column density map  calculated from emission in Herschel imaging, to confirm consistency 
within uncertainties between the three methods.

\subsection{Near-IR extinction maps of AFGL333}
\label{nirakmap}

The dust column density may be mapped by calculating the extinction to background sources.  Our extinction map is calculated
by  dereddening the $H-K_s$ of background stars to the nominal
average intrinsic   $H-K_s$ of field stars.  i.e.,  $A_K$=1.82 $\times$
$(H-K_s)_{obs}-(H-K_s)_{int}$, where, $(H-K_s)_{int}$=0.2 is considered as the average intrinsic color of field stars
(\citealt{allen2008,gutermuth2009,jose2013}).

\begin{figure}[t]
\centering
\includegraphics[scale = 0.56, trim = 15 0 50 40, clip]{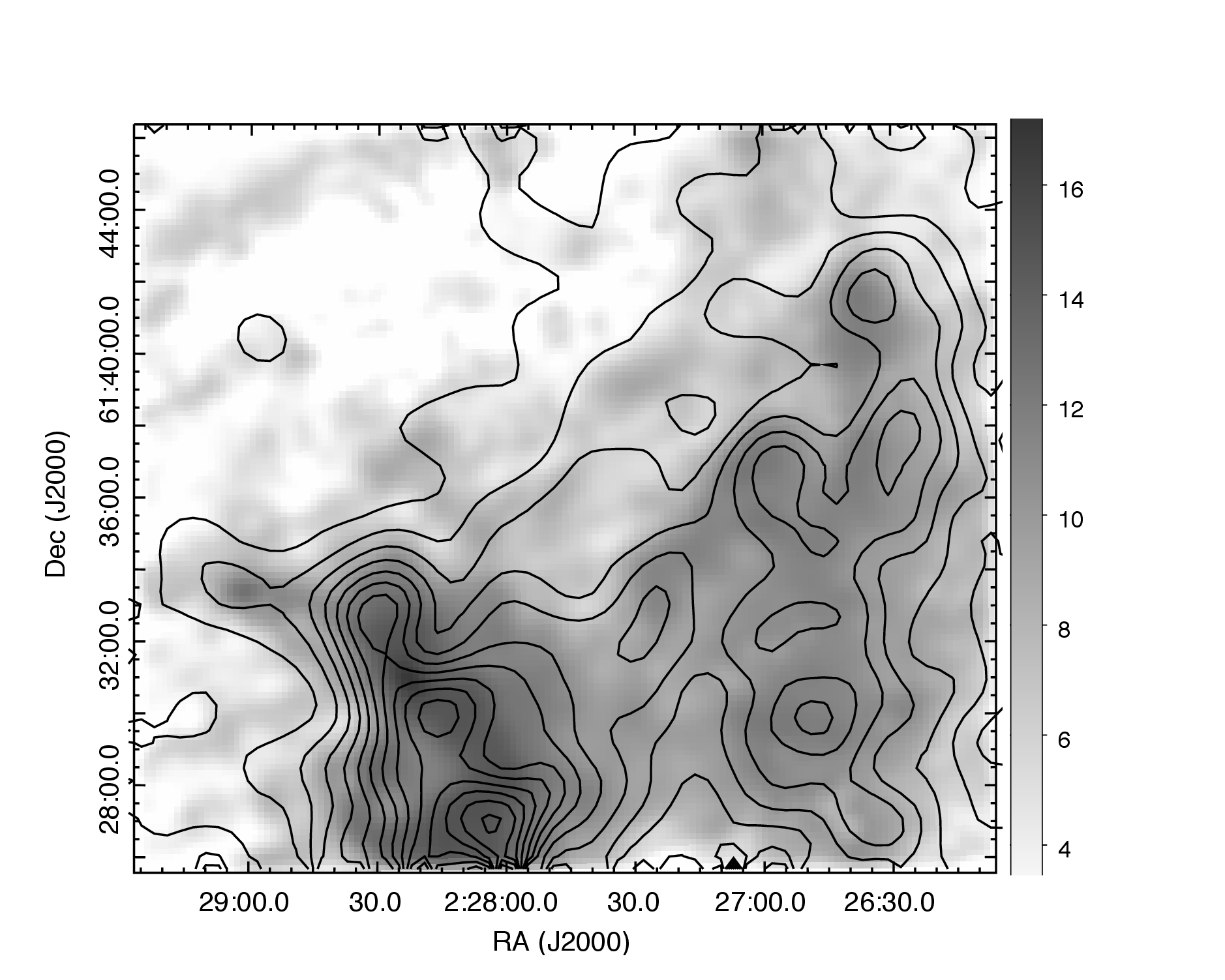}
\caption{ The grey scale represents the $V$-band extinction map made from the  CO  column density map  and the contours
are  the  $V$-band extinction map generated from the  $(H-K_s)$ colors by nearest neighbourhood method assuming $R_V$= 3.1. 
The contours begin at $A_V$=4 mag and increase by 1.5 mag till a maximum of 25 mag. }
\label{extmap}
\end{figure}

Since AFGL333 is situated at a distance of $\sim$ 2.0 kpc, an average foreground extinction of
$A_V$= 2.6 mag  is expected (0.15 mag kpc$^{-1}$ in $K_s$ band, \citealt{indebetouw2005,
chavarria2008}; $A_K$/$A_V$ = 0.114; \citealt{cardelli1989}).   This value is consistent with
the  minimum interstellar  reddening obtained towards  W3  by \citet{oey2005} and also with the 
all-sky dust map based on PAN-STARRS and 2MASS photometry by Green et al. (2015).
Thus in order to eliminate the foreground  contribution, only those stars with $A_V$ $>$ 2.6 mag are used
for getting the  extinction map. To generate the extinction map, the method described by \citet{gutermuth2005} is followed.
Briefly, the region is divided into uniform grids and the  mean and standard deviation of  $A_V$ values of  N nearest neighbour
stars from the center of each grid was measured. The algorithm rejects any stars with  $A_V$ values $>3\sigma$ from the mean.
The  outlier rejection in this application should primarily remove the young stellar objects  with IR excess, where $H-K_s$ excess mainly arising
from their circumstellar disk \citep{meyer1997}.  The resulting $A_V$ map is
convolved with a Gaussian kernel to get the final mean value.  The final extinction map is generated with N=6 and  angular
resolution of $20^{\prime\prime}$ (Fig. \ref{extmap}), after several iteration to achieve a good compromise between resolution
and  noise \citep{gutermuth2009}.

The average  value of $A_V$  across AFGL333 is found to be  $\sim$ 10 mag. The extinction map of  AFGL333 given  in Fig. \ref{extmap}
shows two distinct features: a cavity with low extinction ($A_V$ $<$3.5 mag)  towards the eastern side of the image
and  a highly extincted western half.
A comparison with the sub-mm  and C$^{18}$O(J=1-0) observations (see \citealt{moore2007,sakai2006}) shows that the highest extinction
areas coincide with the dense molecular  cores detected towards AFGL333,  including  the curved features of BRC 5 and AFGL333 Ridge.

The uncertainty in the extinction measurement is dominated by the systematic error in the adopted extinction law, assumed here to be 
based on a total-to-selective extinction $R_V=3.1$ typical of the diffuse interstellar medium.
However, dense clouds usually have  $R_V$ values $>$ 4-5.5, especially at $A_V>20$ mag (\citealt{mathis1990,chapman2009}), 
which would lead to a $\sim 20\%$ overestimate in the extinction (for $R_V$ = 5.5, $A_K$/$A_V$ = 0.134; \citealt{cardelli1989}). 
Only the molecular ridge, which totals $<$ 10\% of the total cloud area, has $A_V$ $>$ 20 mag.

Extinction estimates from stellar colors should be calculated using only background sources.  However, the diskless members of AFGL333
cannot be distinguished from background stars, and their inclusion in our analysis may lead to underestimating the extinction.  Based on
our extinction-corrected comparison of  AFGL333  to  nearby control field (see Section \ref{yso_nir}), the diskless members
of AFGL333 are $\sim$ 30\% of the stars considered here as background sources in regions with $A_V=10$ mag ($\sim$ 45\% for $A_V=20$ mag).   
The inclusion  of diskless  stars therefore moderately underestimates the  extinction throughout the cloud, and is again especially severe in 
high extinction regions along the molecular ridge.

A smaller uncertainty is introduced by our spectral type sensitivity of background stars.  The 90\% completeness limit in the $K_s$-band
is 17 mag (see Section \ref{completeness}), which corresponds to background stars at 2 kpc of K7 for $A_V=0$ mag, K4 for 10 mag, and G4 for 20 mag.  
These differences will lead to an uncertainty of $\sim 2$ mag in $A_V$, with the ridge again most affected.

%%%%%%%%%%%%%%%%%%%%%%%%%%%%%%%%%%%%%%%%%%%%%%%%%%%%%%%%%%%%%%%%%%%%%%%%%%
%%%%%%%%%%%%%%%%%%%%%%%%%%%%%%%%%%%%%%%%%%%%%%%%%%%%%%%%%%%%%%%%%%%%%%%%

\subsection{Extinction map from the CO column density}
\label{coakmap}

 We used the $^{12}$CO (J=2-1) and $^{13}$CO (J=2-1) position-velocity data cubes for the entire W3 region from
\citet{bieging2011} to derive a gas column density map of the area around AFGL333.  The emission of these two CO 
isotopologues was analyzed with a non-LTE statistical equilibrium model grid and a large velocity gradient radiative 
transfer  code (see \citealt{kulesa2005}) to find the CO total column density at each map pixel.  The model incorporates
gas heating dominated by photons via the photoelectric effect together with the observed CO line intensity,
to estimate the UV radiation incident at each point in the map.  The CO abundance and total hydrogen column
density ($N(H)$ = $N(H I)$ $+$ $2N$ (H$_2$)) are calculated following the photodissociation models of \citet{black1987} and
\citet{vandischoeck1988}.  The visual extinction $A_V$, is found using the ratio $N(H)/A_V= 1.8 \times 10^{21}$~cm$^{-2}$mag$^{-1}$
\citep{bohlin1978}.

Figure \ref{extmap} shows the results of this calculation, plotting the CO-derived visual extinction ($A_V$) in grayscale and 
the $K$-band extinction from the NIR point source catalog as contours. The CO-derived column density maps are subject to systematic 
uncertainties in the assumed values for the $^{12}$CO/$^{13}$CO isotopic abundance ratio and the CO/H$_2$ molecular abundance ratio.   For the 
galacto-centric radius of AFGL333, we assume a $^{12}$CO/$^{13}$CO isotopic ratio of 80 with a probable uncertainty of $\sim$ 30\% 
(e.g., \citealt{milam2005}).  We assume a CO/H$_2$ abundance of $10^{-4}$  also with an uncertainty of $\sim$ 30\%.  The observed CO 
lines become saturated for the highest extinction regions ($A_V \geq 20$~mag), and CO may become depleted onto grains at high density 
and low temperature.  Column densities may therefore be underestimated, but this effect should be limited to a small fraction of the 
total area of the region. 
%Considering all the possible systematic uncertainties in our assumptions, the gas column densities and 
%corresponding extinctions, as derived from the CO data, could be in error by as much as a factor of 2.} 
%The systematic uncertainties are different from those that affect the NIR-derived extinction, however, the agreement in the morphology of 
%the extinctions derived by these methods (cf. Fig. 4) helps confirm the overall morphology of the extinction maps. }

\subsection{Comparing extinction maps}
\label{akcomparison}
%%%%%%%%%%%%%%%%%%%%%%%%%%%%%%%%%%%%%%%%%%%%%%%%%%%%%%%%%%%%%%%%%%%%%%%%%%
\begin{figure*}[t]
\centering
\includegraphics[scale = 0.6, trim = 0 50 0 100, clip]{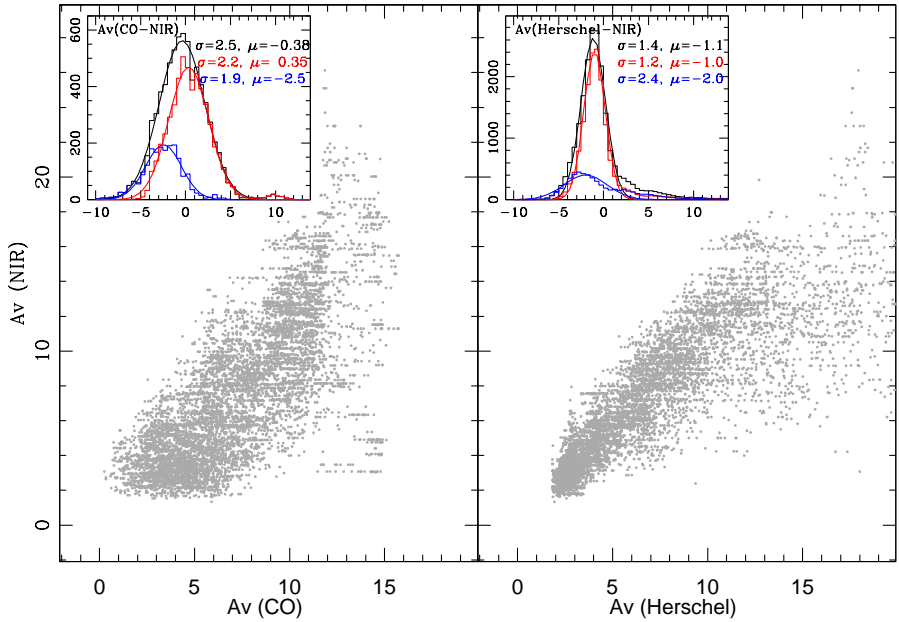}

\caption{Comparison of  $A_V$ measured from near-IR data with CO  derived column density ({\it left}) and with dust column density map calculated from
Herschel far-IR imaging by \citet{ingraham2013} ({\it right}).  The mean and rms of
the gaussian fit to the scattering is shown in the inset plot, where black indicates for all values of  $A_V$, red is for $A_V$ $<$ 10 mag and blue is for
$A_V$ $>$ 10 mag, respectively.     }
\label{correlation}
\end{figure*}
%%%%%%%%%%%%%%%%%%%%%%%%%%%%%%%%%%%%%%%%%%%%%%%%%%%%%%%%%%%%%%%%%%%%%%%%

Figure \ref{extmap} provides qualitative support for the general cloud structure, as calculated from the near-IR extinction
and CO column density.  These results are also qualitatively consistent with the dust column density map calculated from
Herschel far-IR imaging by \citet{ingraham2013}.  However, each of these methods suffers from biases and uncertainties
\citep[e.g.][]{pineda2010}, all of which are difficult to independently evaluate. In this subsection, we compare results 
from different methods to measure extinction to quantify empirical uncertainties in the extinction map.

Figure \ref{correlation} compares the average near-IR extinctions to extinctions estimated from CO column density and 
Herschel dust column density maps within grids of size 38$^{\prime\prime}$.  Below $A_V=10$ mag,  the three sets of extinctions 
are well correlated.  The extinctions from CO column density are 0.35 mag in $A_V$ smaller than those from near-IR colors, 
but with a large scatter of rms$=2.2$ mag between the two values.  The extinctions from the far-IR dust column densities 
are much more tightly correlated with the near-IR extinctions, with an rms of 1.2 mag., but are $1$ mag smaller than
 the near-IR extinction.  At $A_V>10$ mag, these correlations become much worse.  The near-IR colors overestimates 
high extinctions by $2-2.5$ mag, with a scatter of $2-2.5$ mag that is caused by a tail of very high extinction that is undetected in the near-IR.

The tight correlation between the dust emission and near-IR extinction indicates random uncertainties of $\sim 1$ mag in regions 
of low-modest extinction.  The absolute comparisons for both the dust emission and the near-IR extinction depend sensitively 
on correction for foreground and background dust.  The CO column density map should be robust to these uncertainties, and 
establishes that systematic uncertainties are $\sim 0.35$ mag.   The near-IR extinction map becomes much more unreliable 
when $A_V>10$ mag, as expected from our description of uncertainties in this method in Section \ref{nirakmap}.  However, at 
high column densities, these comparisons are sensitive to the filling factor of dense gas and dust
within the $38^{\prime\prime}$ boxes. If the dense material is concentrated in small regions, the near-IR extinction map may be a better
estimate of the median extinction in a region.  The near-IR extinction map has no sensitivity to these optically thick regions, since few
background stars are detectable.

These comparisons establish that the near-IR extinction map therefore provides a reasonable estimate of the extinction to most stars in 
AFGL333 at low extinction.   This map will be adopted for the remainder of the paper.  The extinction map and our stellar census is 
unreliable in the densest regions, which occur mainly along the AFGL333 Ridge.

\section{Census of young stellar objects in AFGL333}
\label{yso}

Several classification schemes have been developed in recent years to differentiate between YSOs with dominant envelope emission,
circumstellar disk emission and diskless sources.
%YSOs emit excess radiation in the infrared %compared with that seen in main-sequence  photospheres,
%due to thermal emission from circumstellar material. %Thus, YSOs can be identified by looking for infrared excess emission.
YSOs are often categorized  into Class 0, I, II or III evolutionary stages \citep{lada1984}. Class 0 objects are deeply embedded
protostars experiencing cloud collapse and are extremely faint at wavelengths shorter than 10 $\mu$m.
Class I YSOs are infrared bright with the emission  dominated by their  spherical envelope. With only near-IR and mid-IR  data it is
impossible to distinguish Class 0 from Class I objects. Hereafter these objects are grouped together and called  Class I in this study.
A Class II YSO has no envelope and is characterised by the presence of an optically
thick, primordial circumstellar disk, which gives excess emission in IR. When  the circumstellar disk  becomes optically
thin, the star is classified as  Class III.

In this section, we identify and classify  Class I and Class II sources using  the 1 - 24 $\mu$m SEDs.
We do not  classify the diskless Class III YSOs because they are indistinguishable from the field stars in
their IR colors (see Section \ref{diskfraction} for details). In the first step, we use those sources detected in all
four IRAC bands to classify the objects as  Class I / Class II based on their color excesses.
Selection of YSOs based on the four IRAC band colors may not be complete, as we are likely to have missed  many sources that fall in the
region with bright nebulosity in [5.8] and [8.0] $\mu$m  bands as well as due to the limited sensitivity  of these two bands (see Table \ref{catalog}).
In order to account for the missing YSOs in [5.8] and [8.0] $\mu$m bands, we identify more YSOs based on their color excess
in $H$, $K_s$, $[3.6]$ and $[4.5]$ $\mu$m. Finally,  we re-examine the entire catalog of sources having  24 $\mu$m photometry
and more YSOs are added to the list based on the color excess of 24 $\mu$m in combination with any IRAC bands.  Possible contaminants are also discussed.
 Estimating the total population of AFGL333 members requires identifying the diskless Class III population.  Since we cannot 
discriminate between foreground and background stars, this estimate relies on statistical estimates of the foreground and 
background populations from a nearby control region.  We consider a control field  region from our near-IR images centered at 
$\alpha_{2000}$ = $02^{h}28^{m}28^{s}$; $\delta_{2000}$ = $+61^{\circ}42^{\prime}45^{\prime\prime}$, $\sim$ 10$^\prime$ 
north of BRC 5. This region is  devoid of nebulosity in $K_s$ band.
The column density map from dust emission  by \citet{ingraham2013} shows that the region  suffers from
normal interstellar reddening. Therefore we consider this a safe region to be used as a control field region.

\subsection{Selection of YSOs using  IRAC data}
\label {yso_irac}

%%%%%%%%%%%%%%%%%%%%%%%%%%%%%%%%%%%%%%%%%%%%%%%%%%%%%%%%%%%%%%%%%%%%%%%%%%
\begin{figure*}[t]
 \centering
\includegraphics[scale = 0.7, trim = 2 40 0 220, clip]{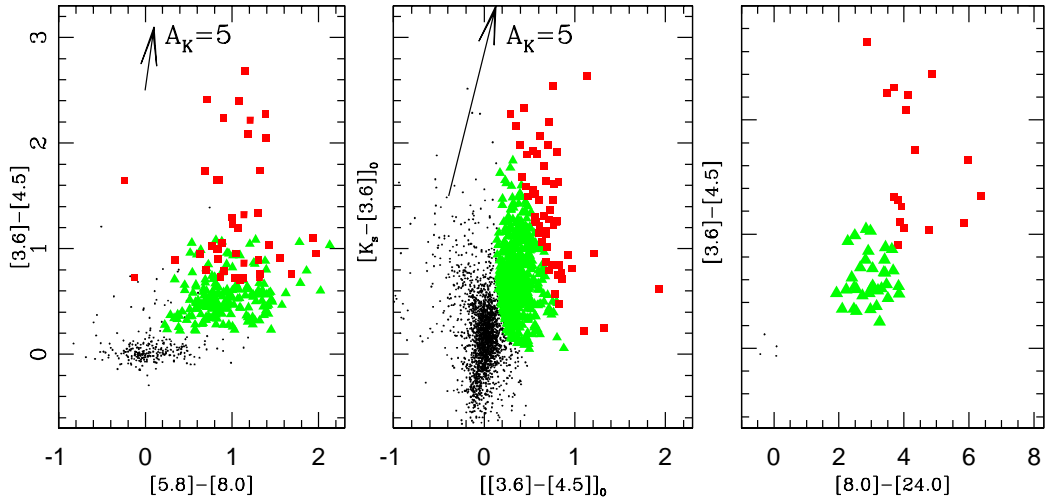}

\caption{Color-color diagrams of all the  sources  (black dots) identified within AFGL333. The YSOs classified as Class I and Class II
based on the color criteria  by \citet{gutermuth2009} are shown using red rectangles and green triangles, respectively.
The reddening vector for $A_{K}$ = 5 mag is plotted using the reddening law from \citet{flaherty2007}. {\it left}: [3.6]-[4.5] 
vs [5.8]-[8.0]  color-color diagram of all the  IRAC sources, {\it middle}: dereddened $[K_s-[3.6]]_0$ vs $[[3.6]-[4.5]]_0$ 
color-color diagram for all the IRAC sources having NIR counterparts. {\it right}: [3.6]-[4.5] vs [8.0]-[24.0] color-color diagram of 
all the  IRAC sources with counterparts in 24$\mu$m.   }

\label{irac-cc}
\end{figure*}
%%%%%%%%%%%%%%%%%%%%%%%%%%%%%%%%%%%%%%%%%%%%%%%%%%%%%%%%%%%%%%%%%%%%%%%%  

 In order to classify the YSOs detected in all IRAC bands into Class I and  Class II, we adopt  the  color criteria given by \citet{gutermuth2009}.
Of the 1567 sources detected in all the IRAC bands (see Section \ref{spitzer}),  538  sources are  within the AFGL333 area 
considered for this study. Of these, 40  have colors consistent with Class I   and 188  have colors consistent
with Class II, respectively. However, in rare cases, highly reddened Class II sources could have  the colors of a Class I source.
In  the highly reddened regions  (i.e., the ridge), the average  $A_V$  $\sim$ 20 mag  can cause a shift of
$\sim$ 0.18 mag in the [3.6]-[4.5] color of a star \citep{flaherty2007}.  If the color excess of Class II YSOs are shifted due to the 
presence of reddening up to $A_V$  $\sim$ 20 mag,    $\sim$ 20\% of Class I YSOs  are expected to be  misclassified Class II sources. 
In Fig. \ref{irac-cc} (left panel), the  [3.6]-[4.5] vs [5.8]-[8.0] color-color diagram is shown, where the Class I and Class II 
sources are shown in red  and green colors, respectively. 

\subsection{YSOs from $H, K_s, [3.6]$ and $[4.5]$ $\mu$m data}
\label{yso_2mass}

Additional YSOs are identified based on their color excess in $H$, $K_s$, 3.6 and 4.5 $\mu$m bands.  We dereddened the individual
point sources based on their location on the extinction map (see Fig. \ref{extmap})   using the extinction law by \citet{flaherty2007}.
We followed the various color   criteria by \citet{gutermuth2009} to classify
the YSOs in the Class I and Class II category. After removing the YSOs which are 
already identified in Section \ref{yso_irac}, a total of 434 more sources are added
to the YSO list. Of these, 30 and 404  sources have colors consistent with Class I and Class II, respectively.  However, the
intrinsic uncertainty of $\sim$ 2 mag in $A_V$ measurement (see Section \ref{akmap}) can cause a variation in the number of
Class I sources by $\sim$ 20 \% and Class II sources by $\sim$ 7\%. Fig. \ref{irac-cc} (middle  panel) shows the dereddened  $K_s$-[3.6] vs
[3.6]-[4.5] color-color diagram, where  Class I and Class II sources are shown in red and green colors, respectively. 

\subsection{Additional YSOs from $H$, $K_s$ and [4.5] $\mu$m data}
\label{yso_nir}

%%%%%%%%%%%%%%%%%%%%%%%%%%%%%%%%%%%%%%%%%%%%%%%%%%%%%%%%%%%%%%%%%%%%%%%%%%
\begin{figure*}
\centering
\includegraphics[scale = 0.8, trim = 2 260 0 0, clip]{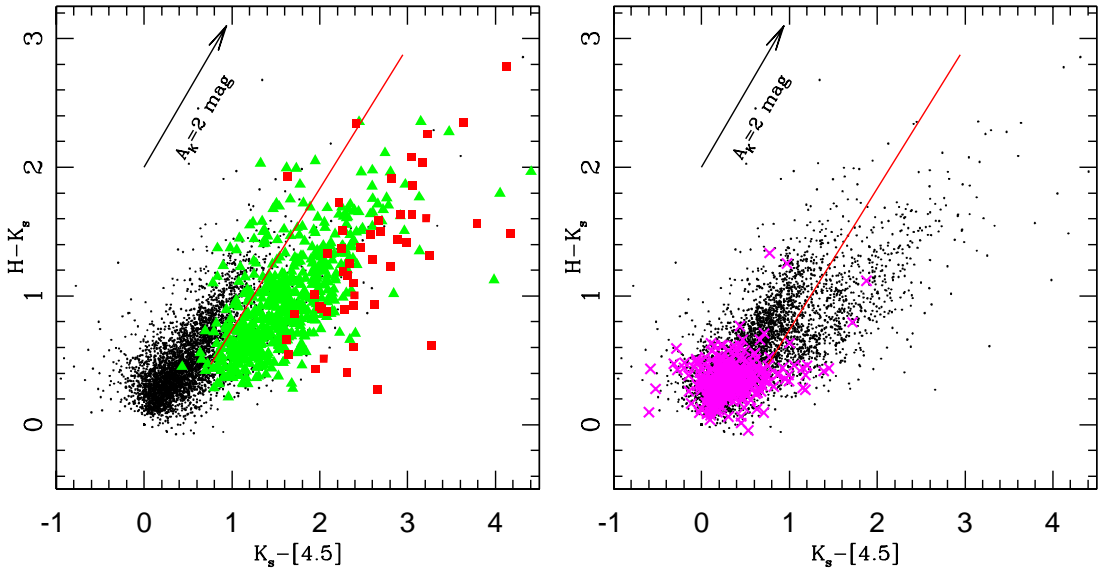}

\caption{{\it  left}: $K_s-[4.5]$ vs $H-K_s$ color-color distribution  of all the  sources (black dots) within AFGL333 along with the  Class I 
(red rectangles) and Class II (green triangles) YSOs identified. Red straight line represents the reddening vector drawn from the tip of 
the dwarf locus  \citep{patten2006} using the reddening law from \citet{flaherty2007}. {\it  right}: Same as in left panel.  Magenta sources represent 
the distribution of sources in the nearby control field. }

\label{irac-cc1}
\end{figure*}
%%%%%%%%%%%%%%%%%%%%%%%%%%%%%%%%%%%%%%%%%%%%%%%%%%%%%%%%%%%%%%%%%%%%%%%%

%Our IR photometry catalog is benifited by the much deeper sensitivity of the $JHK_s$-bands. 
We used a combination of $H,K_s$ and 4.5 $\mu$m bands to identify the additional YSOs within AFGL333 \citep{samal2014}.
Among all the IRAC bands, the 4.5 $\mu$m band not only provides the largest number of point source detections, but also has advantage of  
being unaffected by polycyclic aromatic hydrocarbons (PAHs). Three other  IRAC bands are contaminated by PAHs 
(\citealt{whitney2008,povich2013}).
Fig. \ref{irac-cc1} (left panel) shows the $K_s-[4.5]$ vs  $H-K_s$ color-color distribution of all sources detected
within AFGL333  along with the candidate YSOs identified in the previous sections. The reddening vector from the tip of the
dwarf locus  \citep{patten2006} is also shown. 

 Fig. \ref{irac-cc1} (right panel) shows the color-color distribution of sources in the 
nearby control field region (see setcion \ref{areaofinterest}), which should have the distribution of non-YSO sources in the same Galactic direction 
as that of AFGL333. A comparison of the distribution of YSOs 
already identified in the previous sections (Fig. \ref{irac-cc1},  left panel)
and control field region shows  that all the sources that are located towards the right side of the reddening vector are likely to  have NIR excess. 
After excluding the YSOs that are already identified, several more sources are located towards the right side of the reddening vector. 
Those sources with $H-K_s >$ 0.65 mag and with an excess $>$ 3$\sigma$ 
(where $\sigma$ is the average uncertainty in color) from the reddening vector (see Fig. \ref{irac-cc1}) are considered as YSOs.
 Thus a  total of   121 more  sources are added to the  YSO list.  In order to confirm  these sources as YSOs, we checked their location 
on the $H-K_s$ vs $K_s$ - [3.6] diagram and all the sources
except four, which have  3.6 $\mu$m detection, are found to be  located towards right side of the reddening vector.   
These sources also satisfy the color-color  criteria to identify YSOs using $J$, $H$ and 4.5 $\mu$m data 
given by \citet{zeidler2016} (i.e., $K_s$-[4.5] $>$ 0.49 mag and $J-H$ $>$ 0.7 mag). 
Also, $\sim$ 70\% of these sources satisfy the  color and magnitude selection scheme implemented by \citet{ingraham2011} using  $J,H,K_s$ data.
  
 By selecting   only those objects towards the right side  of the reddening vector and by applying an additional color cut at $H-K_s >$ 0.65, 
we reduce the number of contaminating background objects that are reddened by the clouds. 
A few YSOs may be missing in this approach, but the selected sources are more reliable candidates with NIR excess (\citealt{samal2014,zeidler2016}). 
Since most of these newly identified NIR excess  sources fall in the regime of the already identified Class II YSOs, 
we include them in the Class II category.

\subsection{YSOs using 24 $\mu$m excess}
\label{yso_mips}

In the final step, the entire catalog of  sources with 24 $\mu$m photometry is re-examined.    
Any source that lacks a detection in some IRAC bands but has bright  24 $\mu$m photometry (i.e., [24]$<$ 7 mag; [X]--[24] $>$ 4.5; 
where [X] is any IRAC band) is likely to be a deeply embedded protostar \citep{gutermuth2009}. Using this criteria, we identify 29
new Class I sources. All  previously identified Class I YSOs with MIPS detections  have colors [5.8]--[24] $>$ 4 or [4.5]--[24] $>$ 4
that confirm their classification. The right panel of Fig. \ref{irac-cc} shows the [3.6]-[4.5] vs [8.0]-[24] color-color
diagram of all the  IRAC sources having counterparts in 24 $\mu$m. The candidate  Class I and  Class II   sources are marked in red
squares and green  triangles,   respectively.

\subsection {Final catalog of YSOs}
\label{finalysocatalog}

Our NIR and MIR color criteria  using MIPS, IRAC and NEWFIRM photometry identifies 812 candidate YSOs (99 Class Is and 713 Class IIs) 
associated with the AFGL333.  
Table \ref{ysocatalog} summarizes the statistics of YSOs in various bands. Sample entries of  photometric data of all the stars within AFGL333 is 
given in Table \ref{ysotable} and the full version is available in electronic format.  

 The latest census of YSOs (Class I and Class II) in AFGL333  by \citet{ingraham2011} identified
156 YSOs within  the area considered in this study. Of these, 138 are recovered in our YSO list (37 Class I, 101 Class II).
The slight discrepancy in the number of YSOs is due to the different YSO selection criteria adopted in these two studies. 
Of the 37 Class I sources in our list, 31 are classified as  Class 0/I and 6 as Class II  in \citet{ingraham2011}. 
Of the 101 Class II sources in our list, 91 are classified as Class II and 10 as class I in their list. 
They have also identified an additional 83 candidate pre-main sequence sources in the region using a 2MASS based color-color and
magnitude selection scheme. Of these, 70 are recovered in our YSO list (3 Class I, 67 Class II). 
Their survey was limited by the sensitivity of 2MASS data. In conclusion, the previous survey of this region identified 239 candidate 
YSOs and due to the  improved sensitivity of  $JHK_s$ data in this study
by $>$ 3 mag in each bands, we have identified  573 more YSOs within AFGL333.

%%%%%%%%%%%%%%%%%%%%%%%%%%%%%%%%%%%%%%%%%%%%%%%%%%%%%%%%%%%%%%%%%%%%%%%%%%
\begin{center}
\begin{table}[h]
\caption{YSO catalog summary in various bands}
\label{ysocatalog}
\centering
\begin{tabular}{lcc}
\tableline\tableline
 Band & Class I   & Class II \\
  \tableline

IRAC          & 40 & 188  \\
$H,K_s,3.6,4.5$ & 30 & 404  \\
$[24]+$IRAC(any)& 29 & $-$   \\
$H,K_s, 4.5$         & $-$  & 121 \\
total         & 99 & 713  \\
\tableline
%%\hline
\end{tabular}
%\end{tabular}
\end{table}
\end{center}
%%%%%%%%%%%%%%%%%%%%%%%%%%%%%%%%%%%%%%%%%%%%%%%%%%%%%%%%%%%%%%%%%%%%%%%%%%

\begin{table*}
\centering
\caption{Photometric data$^a$ of all point sources  within AFGL333. }
\label{ysotable}
\begin{center}
\scriptsize
\begin{tabular}{ccccccccccc}
 \hline

 $\alpha_{(2000)}$& $\delta_{(2000)}$ &$J$ & $H$ & $K_s$ & $[3.6]$ & $[4.5]$ & $[5.8]$ & $[8.0]$ & $[24]$ & Class  \\
 deg        &  deg   & mag    & mag       & mag       & mag & mag & mag & mag & mag   \\
\hline

36.9357& +61.6750 &12.10 &11.41 &10.88 & 9.98&  9.74 & 9.38&  9.13& -     & Class II \\
36.6905& +61.6978 &12.16 &11.49 &10.93 & 9.53&  9.19 & 8.40&  7.64& 5.18 & Class II \\
37.1812& +61.4949 &12.32 &11.40 &10.80 &10.26&  9.60 & 8.91&  7.54& 4.15 & Class II  \\
36.6395& +61.6879 &12.68 &11.96 &11.53 &11.13& 10.74 &    - &      -& -	    & Class II  \\
37.1809& +61.5262 &12.81 &12.28 &11.88 &11.21& 10.90 &10.72& 10.53& -     & Class II  \\
36.7891& +61.9122 &12.87 &12.14 &11.72 &11.23& 10.96 &10.67& 10.31& -     & Class II  \\
36.8007& +61.6182 &12.94 &11.51 &10.49 & 9.38&  8.90 & 8.49&  7.53& 4.60 & Class II \\
36.7908& +61.4332 &14.92 &14.18 &13.66 & 13.19& 13.13&12.68& 11.54& -    & Class III/field stars\\

\hline
\end{tabular}\\
$^a$ The complete table is available in electronic form.
\end{center}
\end{table*}

%%%%%%%%%%%%%%%%%%%%%%%%%%%%%%%%%%%%%%%%%%%%%%%%%%%%%%%%%%%%%

\subsection {Source contamination rate}

 Our candidate YSO population may be contaminated by the background sources including  PAH-emitting
galaxies, broad-line AGNs, unresolved knots of shock emission, PAH-emission contaminated apertures etc., that mimic the colors of
YSOs. Since we are observing through the Galactic plane, contamination due to galaxies should be negligible \citep{massi2015}.
In order to have a statistical estimate of possible galaxy contamination in our YSO sample, we used the {\it Spitzer} Wide-area
Infrared Extragalactic (SWIRE) catalog  coming from the observations of the ELAIS N1 field \citep{rowan2013}.  SWIRE is a survey of the
extragalactic field using {\it Spitzer}-IRAC and MIPS bands and can be used  to predict the number of galaxies with colors
that overlap with  YSO colors \citep{evans2009}.   The SWIRE catalog
is resampled for the spatial extent  as well as the sensitivity limits of our observations  in AFGL333 and is also reddened by the average
reddening of AFGL333  (i.e. $A_V$ = 10 mag,  see Section \ref{akmap}). The YSO selection criteria is applied to the resampled SWIRE catalog
and  only $\sim$ 2\% of our YSO sample found to be compatible with the galaxy colors. Similarly, using the color criteria given by
\citet{robitaille2008}, $\sim$ 2\% of YSOs seem to have colors consistent with AGB stars.

Finally, a comparison between the distribution of control field stars and  the YSOs (see Fig. \ref{irac-cc1}) shows that $<$ 5\% of YSOs coincide 
with the location of field stars.   In summary, the contribution of various contaminants in our YSO sample is $<$ 5 \% (i.e., galaxies $\sim$ 2\%, AGBs $\sim$ 2\%),
 which is a small  fraction of the total number of YSOs.

We also applied the   \citet{gutermuth2009} color-color criteria to the candidate YSOs identified within AFGL333. 
$\sim$ 38\% of Class I and $\sim$ 15\% of  Class II of the final YSO  list in Table  \ref{ysocatalog} are matched with the colors of PAH contaminated apertures, 
shock emissions,  PAH emitting galaxies and AGNs.  The major contribution is from the contamination by  AGNs.
However,  several studies  (e.g., \citealt{koenig2008,ingraham2011,willis2013}) have noticed that the use of
\citet{gutermuth2009} criteria  for a region at $\sim$ 2kpc would likely provides an overestimation of the contamination.

\subsection{Mass completeness limit of  YSOs}
\label{masscompleteness}

The  $J$ vs $(J-H)$ and $H$ vs $(H-K_s)$ color-magnitude diagrams (Fig. \ref{jhj}) are used  to get an estimate of the mass
range of the candidate YSOs identified within AFGL333. The Class I and Class II YSOs identified from various color combinations
of IRAC and NIR bands in Section \ref{yso} are shown using red rectangles and green triangles, respectively in Fig. \ref{jhj}.
The pre-main sequence isochrone with  age 2 Myr \citep{bressan2012}  and reddening vectors for several  masses are also
shown. The wide variation in the colors of YSOs in Fig. \ref{jhj} is  caused by  variable extinction and
different evolutionary stages of sources within AFGL333.  In general, the majority of YSOs lie in the mass
range $\sim$ 0.1 - 2 $M_{\odot}$. 

The mass completeness limit for our YSO survey is dictated by the photometric completeness and  wide range of YSO  colors.
We selected the YSO sample   based on the IRAC, $HK_s$+IRAC and IRAC+MIPS   color   combinations (see Section {\ref{yso}).
Within AFGL333,  85\% of the Class I sources have counterparts in
IRAC 4.5 $\mu$m band and all the  Class II sources are detected in 4.5 $\mu$m (see Table \ref{ysocatalog}).
The 90\% photometry completeness limits of $K$ and 4.5 $\mu$m-bands are  estimated to be $\sim$ 17 and 16 mag, respectively
(see Section \ref{completeness}). Assuming a distance of 2.0   kpc for AFGL333 and average extinction $A_V$ as $\sim$ 10 mag (see
Section \ref{akmap}),  the photometric completeness limits in these bands  correspond to an approximate
stellar mass of $\sim$  0.2 M$_\odot$  for a YSO of  age  $\sim$ 2 Myr (using the evolutionary tracks of \citealt{bressan2012}).
The completeness limit is $>$  0.4 M$_\odot$ at high extinction regions such as the AFGL333 Ridge, where the extinction $A_V$ $>$ 20  mag. 
However,  only $<$ 10\% of the total area suffers this high extinction.
Also, due to the  IR excess, the YSOs will be brighter in IR bands compared to their main-sequence counterparts  with only photospheric emission. So the 
completeness limits of Class I and Class II sources will be lower than 0.2 M$_\odot$, when compared to the diskless sources. 
 The previous YSO survey of this region \citep{ingraham2011} was limited by the 2MASS completeness limit, which corresponds to
$>$ 1 M$_\odot$ for $A_V$  $\sim$ 10 mag. With the deep photometry, the current analysis probes the very low mass stars ($>$ 0.2 M$_\odot$) of the region.
For further analysis, we include only those sources brighter than 16 mag in 4.5 $\mu$m.

%%%%%%%%%%%%%%%%%%%%%%%%%%%%%%%%%%%%%%%%%%%%%%%%%%%%%%%%%%%%%%%%%%%%%%%%%%
\begin{figure*}[t]
\centering
\includegraphics[scale =0.75 , trim = 10 100 10 150, clip]{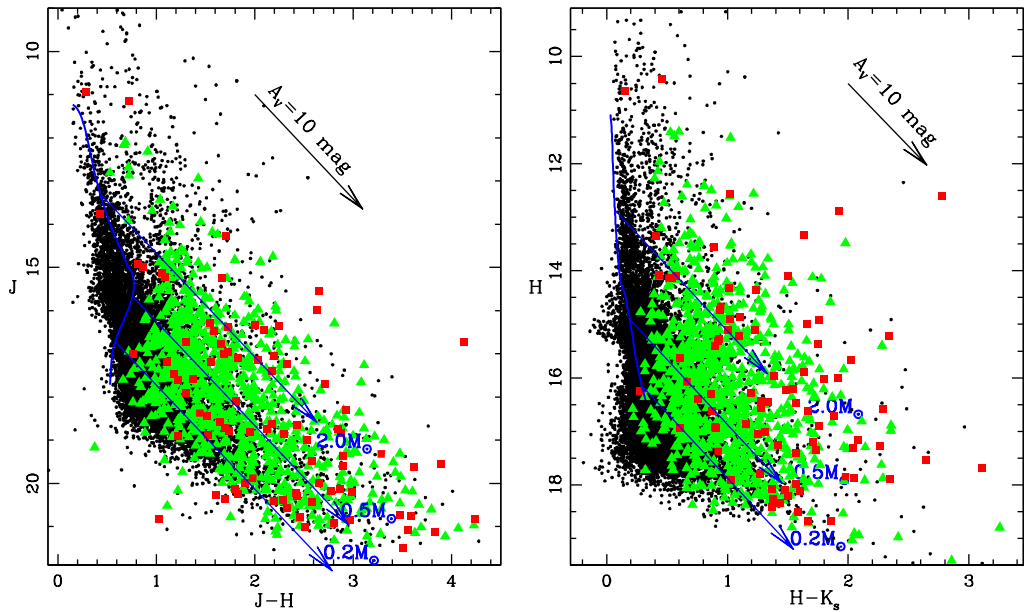}

\caption{ $J$ vs $(J-H)$ ({\it left}) and $H$ vs $(H-K_s)$ ({\it right}) color-magnitude diagrams of all the sources  (black dots)
within AFGL333 region. The red rectangles and green triangles  are the Class I and Class II YSOs identified in
Section \ref{yso}.   Blue solid curve is the pre-main sequence isochrone of age  2Myr \citep{bressan2012} shifted for a
distance of 2.0 kpc. The straight lines represent  the reddening vectors for $A_V$= 20 mag that correspond
to stellar masses of 0.2, 0.5 and 2 $M_{\odot}$.  }

\label{jhj}
\end{figure*}
%%%%%%%%%%%%%%%%%%%%%%%%%%%%%%%%%%%%%%%%%%%%%%%%%%%%%%%%%%%%%%%%%%%%%%%%

\section{Characteristics of young stellar objects, environment  and star formation in AFGL333}

In this section  the final YSO catalog obtained from Section \ref{yso} is used to obtain their spatial
distribution, mass distribution, stellar surface density and  clustering properties.

\subsection{Spatial distribution of YSOs and sub-clustering in AFGL333 } 
\label{spatial}

The spatial distribution of YSOs can probe the fragmentation processes that lead to the formation of protostellar cores
and the subsequent dynamical evolution of star forming regions. Fig. \ref{sdensity} shows  the spatial distribution
of the  candidate YSOs   (i.e., red: Class I; green: Class II;) overlaid  on the  4.5 $\mu$m {\it Spitzer} image,
where the  YSOs are preferentially located in/around the regions of high extinction
(see Fig. \ref{extmap}).  The Class I YSOs follow the elongated nature of the  molecular ridge seen in the CO
maps of \citet{sakai2006} and \citet{bieging2011} and in our extinction map shown in Fig. \ref{extmap}. 

Since W3 is a complex star forming region, a high level of sub-clustering is expected and is evident in the spatial distribution of
candidate YSOs  (Fig. \ref{sdensity}, left panel). In order to understand the level of sub-clustering within AFGL333, we generated the stellar 
density map using the nearest neighbourhood technique \citep{gutermuth2005}. We followed the method introduced by \citet{casertano1985}, where the stellar
density $\sigma(i,j)$ inside a cell of a uniform  grid with center at the coordinates $(i,j)$ is

$\sigma(i,j)$ = $N-1 \over \pi r^2_N(i,j)$

where, $r_N$ is the distance from the center of the cell to the N$^{th}$ nearest source. 
 The value of N is allowed to vary depending on the   smallest scale structures of the regions of interest.
The surface density map for  the candidate YSOs detected above the completeness limit is generated with a  grid size of
10$^{\prime\prime}$  $\times$ 10$^{\prime\prime}$ and N=6 as a compromise between resolution and sensitivity of the map.

In the surface density map shown in  the right panel of Fig. \ref{sdensity},  three major groups of YSOs  are identified,
which  agrees with our visual interpretation of the spatial distribution of YSOs.
The main over-density corresponds to the  group of YSOs  associated with the AFGL333 Ridge and its surroundings, which itself appears 
to have multiple density peaks. Hereafter we name this group of YSOs as AFGL333-main. 
Within AFGL333-main, the YSO surface density  peaks at the center of the cluster associated with the \hii region
(i.e., IRAS 02245+6115; see Fig. \ref{area}).  The second over-density of YSOs is located  towards
the north of AFGL333 Ridge, ($\alpha_{2000}$ = $02^{h}27^{m}14^{s}$;  $\delta_{2000}$ = $+61^{\circ}36^{\prime}48^{\prime\prime}$)
and the YSOs are associated with a  mid-IR cavity seen in the  background image.  The majority of  YSOs in this cluster
are Class II sources. Hereafter we name this group of YSOs as AFGL333-NW1. 
One more cluster (AFG 333-NW2)  is seen further  north west of the region ($\alpha_{2000}$ =
$02^{h}26^{m}34^{s}$;  $\delta_{2000}$ = $+61^{\circ}41^{\prime}27^{\prime\prime}$) and  $\sim$ 70 YSOs are associated with it.
 
 Some of these groups of YSOs are part of known clusters in the literature. For example, the sub-group associated with BRC5, located at 
the outermost boundary of the HDL  (IRAS 02252+6120) and with the \hii region  (IRAS 02245+6115), host clusters
\citep{bica2003}. The latest clustering analysis of the W3 complex using minimum spanning tree method by Rivera-Ingraham et al. (2011) 
identified 8 stellar groups (IDs 0 to 7) as part of AFGL333. Of these, 5 groups (IDs 1, 3, 4, 6 and 7) are within the area considered in this study 
and others are outside the area. The groups 6 and 4 are associated with AFGL333-NW1 and groups 1, 3 and 7  belong to AFGL333-main. 
The individual stellar peaks in Fig. \ref{sdensity} coincide  with that of the stellar groups identified by Rivera-Ingraham et al. (2011).  
Though the individual groups are assumed to be  at various evolutionary stages and at  different
environments (see \citealt{ingraham2011} for details),  due to our limited spatial resolution and detection sensitivity
in this study we are unable to segregate the various sub-groups within AFGL333-main and hence tentatively consider  as a single group of YSOs.

Based on the YSO density distribution, the  extent of individual groups within AFGL333 is estimated from the  
radius  of the outermost stellar density contour (see Fig. \ref{sdensity}) around each cluster, within which $\sim$ 90\%
of the candidate YSOs are concentrated.  Thus radii of 6.0$^\prime$ (3.5 pc), 3.5$^\prime$ (2.0 pc) and 3.0$^\prime$ (1.7 pc) are considered for
AFGL333-main, AFGL333-NW1 and AFGL333-NW2, respectively.  Radii  of  the individual groups are marked in Fig. \ref{sdensity}.
For AFGL333-NW1 and AFGL333-NW2,  stellar density peaks are considered as their center, whereas AFGL333-main has
multiple density peaks and hence the center of the circle which covers the outermost density contour is tentatively considered as its center.

%%%%%%%%%%%%%%%%%%%%%%%%%%%%%%%%%%%%%%%%%%%%%%%%%%%%%%%%%%%%%%%%%%%%%%%%%%
\begin{figure*}
\centering
\includegraphics[scale = 0.51, trim = 10 0 20 0, clip]{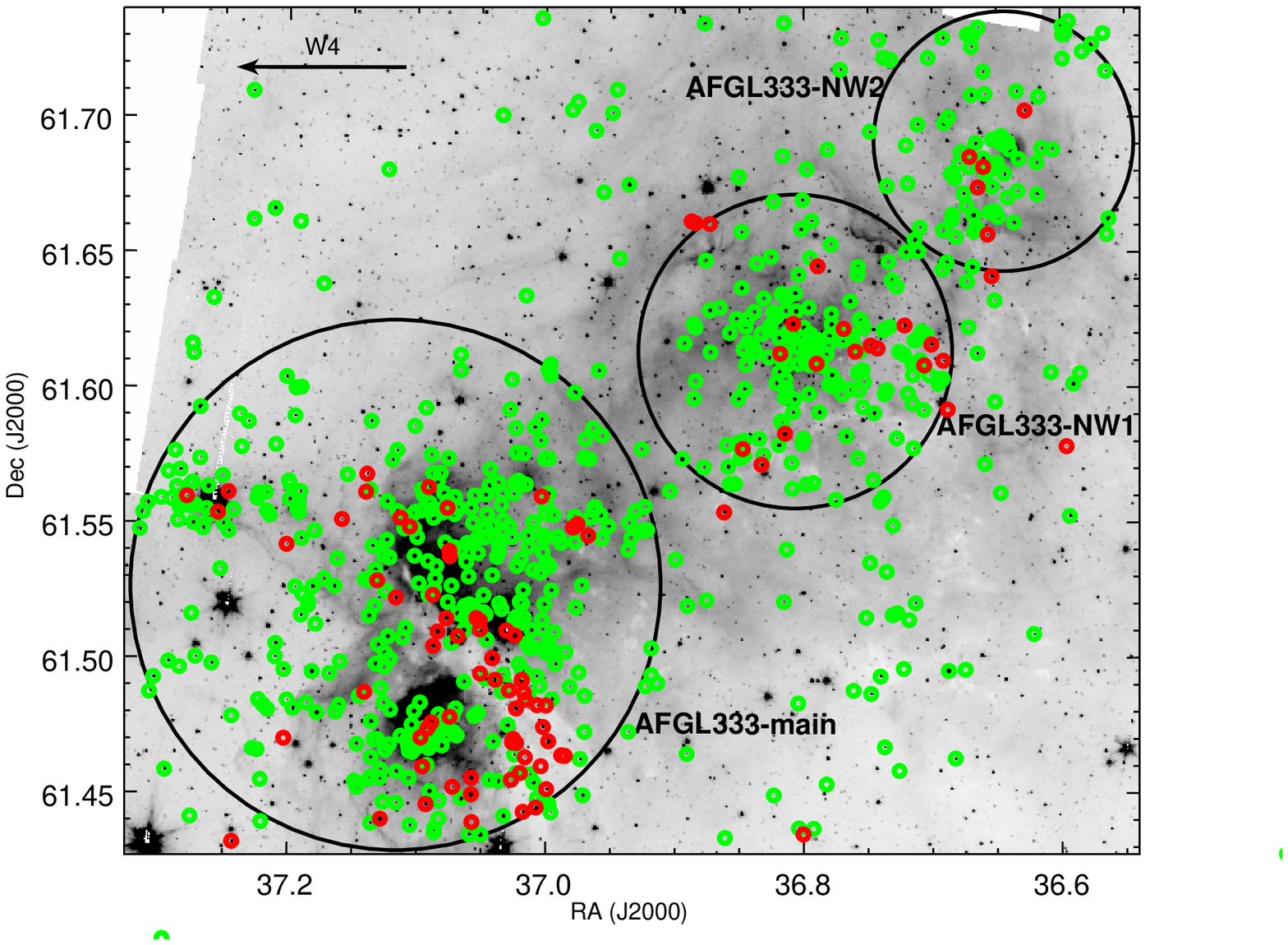}
\includegraphics[scale = 0.28, trim = 0 0 0 0, clip]{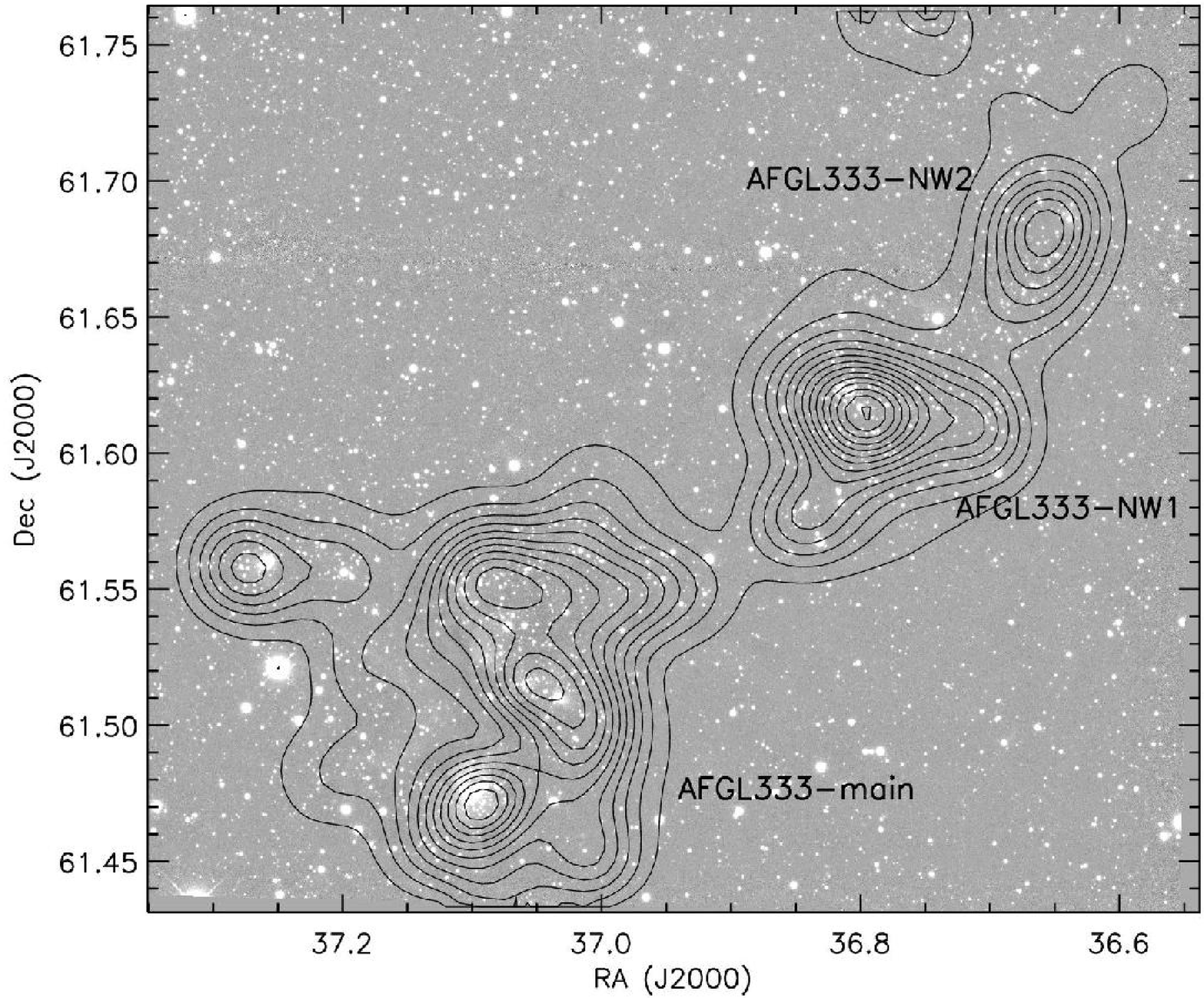}
\caption{{\it left}: Spatial distribution of  candidate YSOs identified within AFGL333   overlayed on the
4.5 $\mu$m  {\it Spitzer}-IRAC  image.  Class I, Class II YSOs are represented in red and green, respectively.
The important stellar groups identified within AFGL333 are labeled and the circles around them represent their
estimated sizes. {\it right}: Surface density map of  YSOs  within AFGL333  overlayed on the NEWFIRM $K_s$-band image.
Surface density was calculated using the 6 nearest neighbours. Contours correspond to YSO surface densities of 5 to 44 stars
pc$^{-2}$ with an interval of 3 stars pc$^{-2}$. }

\label{sdensity}
\end{figure*}
%%%%%%%%%%%%%%%%%%%%%%%%%%%%%%%%%%%%%%%%%%%%%%%%%%%%%%%%%%%%%%%%%%%%%%%%

\subsection{Properties of setllar groups in AFGL333: Disk fraction and Ages} 
\label{diskfraction}

In this subsection we discuss the  disk fraction and  ages of three stellar groups identified within AFGL333. Though these individual groups 
have sub-clusters in it (Fig. \ref{sdensity}), due to the limited sensitivity and spatial resolution of the present study,
we consider them as  single group of YSOs. Hence  the properties of the individual groups studied below should be considered as their average
properties.   The statistics of various 
sources  within individual groups as well as in a nearby control field region are summarized in Table \ref{clusterpars}. 

%%%%%%%%%%%%%%%%%%%%%%%%%%%%%%%%%%%%%%%%%%%%%%%%%%%%%%%%%%%%%%%%%%%%%%%%%%
\begin{center}
\begin{table*}[t]
\caption{Derived cluster parameters}
\label{clusterpars}
\centering
%\begin{tabular}{|p{.40in}|p{.45in}|p{.58in}|p{.28in}|}
\begin{tabular}{lccccccc}
%\hline
\tableline\tableline
 Group  &  Radius & Total & Class I   & Class II & N(I/II)  &disk fraction  \\
  ID &    &   stars &  &     &   &  ($A_V$ corrected)  \\
  \tableline

AFGL333-main & 6.0$^{\prime}$ (3.5pc)   & 1647 & 58 & 372& 0.16& 61\% &   \\
AFGL333-NW1    & 3.5$^{\prime}$ (2.0pc) & 615  & 14 & 158 & 0.09 & 58\% & \\
AFGL333-NW2   & 3.0$^{\prime}$ (1.7pc) &  378  & 5 & 66 & 0.08& 50\%\\
Control field & 3.0$^{\prime}$  (1.7pc) & 279  & - & -     &  -  & & \\

\tableline
%%\hline
\end{tabular}
%\end{tabular}
\end{table*}
\end{center}
%%%%%%%%%%%%%%%%%%%%%%%%%%%%%%%%%%%%%%%%%%%%%%%%%%%%%%%%%%%%%%%%%%%%%%%%%%

An approximate age estimation of individual groups is difficult from the color magnitude diagram because 
the luminosity scatter is large due to NIR excess and extinction (eg. Fig. \ref{jhj}). An alternate  way to roughly infer the ages of 
young clusters is from the disk fraction since the fraction of sources surrounded by circumstellar disks strongly depends on the age of the system
(\citealt{haisch2001,hernandez2008,fang2013}). 
From {\it Spitzer}-IRAC studies of young clusters, the average disk fractions at different cluster ages are 75\% at $\sim$ 1Myr, 50\% at $\sim$ 2Myr,
20 \% at $\sim$ 5Myr and 5\% at $\sim$ 10 Myr \citep{williams2011}. 

The census of the diskless Class III sources is unknown in this survey because of their lack of  IR-excess,and  X-ray or spectroscopic data 
are not available to identify them. In order to account for the total number of Class III sources, the statistics of  the point sources in a nearby
control field is used (Table \ref{clusterpars}). We subtracted the field star contribution from the individual groups, after scaling the 
number of sources in the control field to the areas of individual groups.  The remaining number of sources is considered as the total 
member stars (disk and diskless)  within each region. However, a direct subtraction of  field stars would overestimate the disk fraction
because the background sources within the sub-regions are seen through
the intervening molecular cloud, whereas the control field  stars are only affected by the normal interstellar reddening.
To account for differential extinction between AFGL333 and our control field, we synthesized the background stellar population
using the  Besan\c con  Galactic model of stellar population \citep{robin2003}, for  normal interstellar reddening and for  $A_V$= 10 mag.
Here we assume $A_V$= 10 mag  as the average reddening within AFGL333  (see Section \ref{akmap}).
We find that  $\sim$  85\% of field stars will be seen through the cloud in $L$-band when compared to the total
number of field stars in a cloud free region.

 Table \ref{clusterpars} provides the disk fraction of individual clusters. 
After accounting for   extinction,  the disk fraction is estimated to be  $\sim$ 61\%, 58\% and 50\% for
AFGL333-main, AFGL333-NW1 and AFGL333-NW2, respectively.  The average disk fraction vs age trend reported in the literature
(e.g. \citealt{haisch2001,hernandez2008,fang2012,fang2013}) suggests $\sim$ 50\% of the stars in a given region should have lost their disks around  
2-3 Myr.  The  disk fraction estimated for  AFGL333 matches with star forming regions of age   $\sim$ 2-3 Myr.  

Another  way to infer the relative ages  of different cluster populations is from the Class I/II ratio, given the different
median lifetimes  for these two evolutionary phases (i.e, 0.44 Myr for Class I; 2 Myr for Class II; \citealt{evans2009}). 
The  ratio of Class I to Class II sources should decrease with the average age of the region (\citealt{beerer2010,myers2012}).
Within the sub-clusters of AFGL333,  the class I/II ratios are  $\sim$ 10\% (see Table \ref{clusterpars}), indicating a low
fraction of embedded stellar population.
The Class I/II value for AFGL333 is  comparable to the Vul OB1 \citep{billot2010} and Cygnus X North \citep{beerer2010},
which are at similar distances  to AFGL333.  After scaling to the distance of AFGL333,
the Class I/II ratio of some of the c2d clouds \citep{evans2009} and young  clusters   \citep{gutermuth2009} agrees with that of AFGL333-main
\citep{kim2015}. Assuming the same distance,  all of the above regions and AFGL333 have similar Class I/II values, which indicates
that they are similar in age ($\sim$ 2-3 Myr) or at a similar evolutionary stage. 

As evident in Fig. \ref{sdensity}, the number of Class I sources increases towards the center of the ridge. The Class I/II ratio
($\sim$ 1.3; 32 Class I, 24 Class II) within 2 pc$^2$ of ridge is high when compared to $\sim$ 0.08 in the rest of the cloud.  Therefore  the 
ridge is likely  young ($<$ 1 Myr, \citealt{evans2009,samal2015}) while other regions within AFGL333 are older.  This is in agreement with 
\citet{ingraham2011}, who found that the  stellar group associated with the molecular ridge is relatively younger than that of other stellar groups.

 The above methods to estimate the age of the region should be taken with caution.  The disk fraction can only be used as a 
rough age indicator because of large scatter in this relation due to several factors including the incompleteness of different surveys,  
differences in diagnostic methods used for disk fraction estimation and the effects that age spread, metallicity, UV radiation and 
stellar density may have on the  disk evolution (\citealt{soderblom2014,spezzi2015}).  
Similarly, the  Class I/II ratio is sensitive to the age spreads and  also due to the uncertainty in the Class 0/I lifetimes. Moreover the disk fraction
or Class I/II ratios of AFGL333   are the average values of the individual groups, which  contains several sub-groups and are probably at  various 
evolutionary stages.

\subsection{Estimation of cloud mass}
\label{cloudstructure}

In Section \ref{akmap}, we constructed an extinction map of resolution 20$^{\prime\prime}$ for the AFGL333 region using the deep 
NIR data. From the extinction map and assuming a distance of 2.0 kpc,   the
cloud mass of  AFGL333 is estimated using following relation {\it {M = $m_H$ $\mu_{H_2}$ $A$ $\Sigma$ $N(H_2) $}},
where $m_H$ is the mass of hydrogen,  $\mu_{H_2}$ is the mean molecular weight per $H_2$ molecule, $N(H_2)$ is the column density
and A is the pixel area.  The value of  $\mu_{H_2}$ used is 2.8  \citep{kauffmann2008}.
We convert the extinction in magnitude to column density using the relation $N(H)$/$A_V$ = 1.87 $\times$
10$^{21}$ cm$^{-2}$mag$^{-1}$  \citep{bohlin1978}. This expression has been derived assuming $R_V$ = 3.1.
After subtracting the foreground reddening, the column density of  all   pixels within AFGL333 is integrated to estimate the cloud  mass. 
The  estimated mass within  AFGL333   area is $\sim$ 1.7 $\times$ 10$^4$ $M_\odot$  which is consistent with the mass 
($\sim$ 1.8 $\times$ 10$^4$ $M_\odot$) estimated from the  CO based column density map.  We explain below the various uncertainties 
involved in the mass estimation from the NIR based extinction map and infer a corrected mass.  

The major uncertainty in the estimation of cloud mass comes from the assumption of standard extinction law in the complex. 
Use of $R_V$ = 3.1 would overestimate the cloud mass by a factor of 1.4 when compared to  $R_V$ = 5.5  (i.e., $N_H$/$A_V$ = 1.37 $\times$
10$^{21}$ cm$^{-2}$mag$^{-1}$,  \citealt{heiderman2010,evans2009,draine2003}).   
Similarly, there is an intrinsic uncertainty up to  $\sim$ 2 mag in every pixel of $A_V$ map (see Section \ref{akmap}).
Also, the  estimated cloud mass  should be considered as a lower limit, since the  measured column density
from extinction map is limited by the detection limit of our NIR observations (i.e., $A_V$ $<$ 26 mag), especially towards the
ridge, where the column density is high.

A small fraction of the total area of AFGL333 (i.e., $\sim$ 4 \%), close to the center of the ridge,
has an   average column density $\sim$  3.7 $\times$ 10$^{22}$ cm$^{-2}$ \citep{higuchi2013}. 
The mass of the molecular clump around the ridge was measured as $\sim$ 4.2 $\times$ 10$^3$ M$_{\odot}$ by \citet{higuchi2013} using
C$^{18}$O(J=1-0) line emission. 
 Including the  extra cloud mass which is underestimated in our extinction map,
the total mass of the AFGL333 region is estimated to be  $\sim$ 2.1  $\times$ 10$^4$ $M_\odot$.   The intrinsic uncertainty 
of $\sim$ 2 mag in $A_V$ measurement (see Section \ref{akmap}) is added to the extinction map in order to estimate the upper limit of the  
cloud mass, which is obtained as $\sim$  2.6 $\times$ 10$^4$ $M_\odot$. 
The cloud mass estimate of AFGL333 per area is consistent with that  of \citet{ingraham2013}. 
AFGL333 region contains  $\sim$ 10\%  of the total mass of the entire W3  complex measured from $^{12}$CO(J=3-2), $^{13}$CO(J=3-2)
and C$^{18}$O(J=3-2)  maps 
($\sim$ 4.4 $\times$ 10$^5$ $M_\odot$; \citealt{polychroni2012}) and from the {\it   Herschel} Far-IR dust emission map 
($\sim$ 2.3 $\times$ 10$^5$ $M_\odot$; \citealt{ingraham2013}).

We also  calculated the cloud masses around  individual stellar groups  by integrating the cloud density within their
respective area, after   accounting for the  cloud mass towards the ridge and  the intrinsic  uncertainty in extinction measurements
(see Table \ref{clusterpars2}).

\subsection{Star formation efficiency and rate}

Star formation efficiency (SFE) and rate are two fundamental physical parameters to characterize how the cloud mass is converted into  stars.
We derive  the star formation efficiency of the region  by comparing the mass of the gas reservoir
($M_{cloud}$), with the mass that has turned into stars ($M_{star}$) during the last few million years.
i.e., SFE = $M_{star}$/$(M_{star} + M_{cloud}$) \citep{myers1986}. 

In order to estimate the stellar mass, the  total number of stars detected within individual stellar groups
and brighter than  the completeness limit ($K_s< $ 17 mag, for $>$ 0.2 M$_{\odot}$ at $A_V<$ 10 mag) is used. 
We subtracted the extinction corrected field star contribution from individual groups (see Section \ref{diskfraction}). 
The  remaining stars are normalized to  the Kroupa IMF with $\alpha$ = 2.3 for $M$ $>$ 0.5 $M_{\odot}$ and $\alpha$ = 1.3 for 
$M$ $<$ 0.5 $M_{\odot}$ \citep{kroupa2002}.  The total stellar mass is derived by extrapolating down to hydrogen burning limit (0.08 $M_{\odot}$). 
The estimated stellar mass within each stellar groups and total stellar mass of three groups  are given in Table \ref{clusterpars2}. 

The overall star formation efficiency  for various stellar groups ranges  between $\sim$ 3\% - 11\% . Their combined
star formation efficiency  is $\sim$ 4.5\%. In general, the star formation efficiency  of AFGL333 is
consistent with that of  majority of the  nearby star forming regions in the Galaxy  (i.e., 3--6\%, \citealt{evans2009}). 

%%%%%%%%%%%%%%%%%%%%%%%%%%%%%%%%%%%%%%%%%%%%%%%%%%%%%%%%%%%%%%%%%%%%%%%%%%
\begin{table*}
\begin{center}
\centering
\caption{Cluster parameters}
\label{clusterpars2}

%\begin{tabular}{|p{.40in}|p{.45in}|p{.58in}|p{.28in}|}
\begin{tabular}{llllllll}
%\hline
\tableline\tableline
 Group  &  area &  $M_{cloud}^a$ & $M_{star}$ &  SFE & SFR$^b$ &$\Sigma_{gas}$ & $\Sigma_{SFR}$ \\
  ID &   (pc$^2$) &  ($M_{\odot}$) &  ($M_{\odot}$)&    & ($M_{\odot}$ Myr$^{-1}$)& ($M_{\odot}$ pc$^{-2}$)   & ($M_{\odot}$ Myr$^{-1}$pc$^{-2}$)  \\
  \tableline

AFGL333-main & 38.0 & 7100 - 7700 & 248 & 0.03 - 0.03 & 78 - 118  & 187 - 203 & 2.1 - 3.1 \\
AFGL333-NW1    & 12.8 & 940 - 1150  & 123 & 0.10 - 0.12 &  30 - 45   & 73 - 90   &  2.3 - 3.5 \\
AFGL333-NW2   & 9.5  & 660 - 810   & 69  & 0.08 - 0.10 &   20 - 31  & 69 - 85   & 2.2 - 3.3   \\
Total        & 60.3 & 8700 - 9700 & 440 & 0.04 -  0.05  &  128 - 192 & 144 - 161 & 2.1 - 3.2 \\

\tableline
%%\hline
\end{tabular}
%\end{tabular}
\end{center}
$^a$ The cloud mass is estimated after incorporating the uncertainty in extinction measurement\\
$^b$  SFR is calculated for ages 2 and 3 Myr \\
\end{table*}
%%%%%%%%%%%%%%%%%%%%%%%%%%%%%%%%%%%%%%%%%%%%%%%%%%%%%%%%%%%%%%%%%%%%%%%%%%

Star formation rate (SFR) is the rate at which the gas in a cloud turns into stars. SFR can be  estimated using the relation,
SFR = $M_{cloud} \times$ SFE/$t_{SF}$, where $t_{SF}$ is the  duration of star formation. The  star formation rates of
the individual stellar groups as well as for the entire  region are estimated by assuming the duration of star 
formation in AFGL333 is $\sim$ 2 - 3 Myr (see Table  \ref{clusterpars2}).  
On average, the AFGL333  region is forming $\sim$ 130 - 190 M$_\odot$ of stars  every Myr. 
The star formation rate varies with gas density ($\Sigma_{gas}$), i.e., the low density regions have star formation
rates less than the high density regions within AFGL333 (see Table \ref{clusterpars2}). This result agrees with
the analysis of  nearby molecular clouds (\citealt{lada2010,gutermuth2011}), {\it Spitzer}-c2d/GB clouds \citep{heiderman2010} 
and  galactic massive dense clumps \citep{wu2010}, where they find a linear relation between star formation rate  and 
local gas density. 

The star formation rate  per unit surface area ($\Sigma_{SFR}$), i.e. the density of star formation, has been considered  as a more generalized 
representation  of the star formation rate of a given region \citep{heiderman2010}. $\Sigma_{SFR}$ for   individual groups within AFGL333  has been 
estimated by considering the  area of each cluster over which the cloud mass is measured (see Table \ref{clusterpars2}). 
The derived values of  $\Sigma_{SFR}$  for individual groups  and the total  $\Sigma_{SFR}$  for the entire region
as well as the corresponding cloud gas surface density ($\Sigma_{gas}$) are listed in Table \ref{clusterpars2}.
For AFGL333,  the average values of  $\Sigma_{gas}$  and $\Sigma_{SFR}$ are $\sim$ 140 - 160 $M_{\odot}$ pc$^{-2}$  and
 $\sim$ 2 - 3 $M_{\odot}$ Myr$^{-1}$pc$^{-2}$, respectively.  The above estimates of stellar  mass and various parameters are 
subjected to   bias introduced by the unresolved binary/multiple systems.
About 75\% of the stars that make up the standard IMF are M type stars which have a binary fraction of 
$\sim$ 20-40\% (\citealt{lada2006,basri2006}). Assuming $\sim$ 30\% of member stars in AFGL333 are unresolved binaries,  the actual number of forming stars,
the star formation efficiency and star formation rates  would be increased by about the same factor \citep{evans2009}.

\section{Discussion}

The W3 giant molecular cloud complex has long been discussed as a classic example of induced or triggered star formation
(e.g., \citealt{lada1978,oey2005}). The expansion of the nearby W4 super bubble may have  
created the `high density layer (HDL)' of molecular gas, at its western periphery. W3 Main, W3 (OH) and AFGL333 are the three main star forming
sites identified within the HDL. W3 Main and W3 (OH) are located at the
edges of the cavity created by the young cluster IC 1795. It has been proposed that the star formation within W3 Main and W3 (OH) are 
induced by IC 1795 (e.g., \citealt{oey2005,zuniga2015}). 
According to \citet{ingraham2013}, in W3 Main the triggering processes work at local (sub-parsec) scales, with high mass
stars acting to confine and compress material, enhancing the efficiency for the formation of new high
mass stars by making convergent flow.
In converging flows, the formation of new stars in cluster can be
facilitated for an extended period as the structures will continue drain matter until the matter in their reservoirs get depleted. 
As pursued by \citet{ingraham2013,ingraham2015}, the combined effects of constructive feedback and converging flow could lead to 
the unique population of high-mass stars and clusters in W3 Main. 
W3 Main and AFGL333 are possibly different in terms of the cloud density and dynamics. W3 Main perhaps had initial high 
density which resulted to become very high mass stars in it  whereas AFGL333 does not have density as high as W3 Main. 

Using deep NIR and {\it Spitzer} data sets, 
we obtained the census of young stellar objects within AFGL333  as well as its cloud structure and mass. In this section,
we compare the star formation rate density and  gas density estimated for  AFGL333 with  nearby low mass clouds as well as high mass
regions. We also compare these parameters between W3 Main and AFGL333, and discuss whether  the differences between them  have any  implication
due to the various star formation scenarios proposed in these two regions. 

\subsection{Comparison of AFGL333 with  other star forming regions and W3 Main}

The surface density of star formation rate ($\Sigma_{SFR}$) ranges from   0.1 and 3.4 M$_\odot$ Myr$^{-1}$ pc$^{-2}$ 
for a sample of 20 local low mass star forming regions \citep{heiderman2010}.
Similarly,  \citet{evans2009} reported the values between  0.6 - 3.2 M$_\odot$ Myr$^{-1}$ pc$^{-2}$, for the nearby 
low mass {\it Spitzer}-c2d/GB clouds.  Both of those studies have used the NIR extinction map for cloud mass estimation, 
similar to the method followed in this study. The $\Sigma_{SFR}$ measured towards AFGL333  ($\sim$ 2 - 3 $M_{\odot}$ 
Myr$^{-1}$pc$^{-2}$; table \ref{clusterpars2})  is comparable to  that  of these low mass regions. 

We also compare the $\Sigma_{SFR}$ of AFGL333 with the  high mass star forming regions such as  NGC 6334, W43 and IRDC G035.39-00.33 
(\citealt{willis2013,motte2012}).  NGC 6334 is located at a similar distance ($\sim$ 1.6 kpc) and the  YSO identification and  
cloud mass estimation  from extinction map  are calculated  in a  similar method as in AFGL333.  $\Sigma_{SFR}$ of 
NGC 6334 is of $\sim$ 13.0  M$_\odot$ Myr$^{-1}$ pc$^{-2}$, factor of $\sim$ 6  higher than  AFGL333. 
For W43 and IRDC G035.39-00.33,  $\Sigma_{SFR}$ is as high as $\sim$ 10-100 M$_\odot$ Myr$^{-1}$ pc$^{-2}$ (\citealt{motte2003,motte2012}). 

W3 Main and AFGL333 are part of the same cloud complex and have similar sizes ($\sim$ 0.45 deg$^2$, \citealt{lada1978,ingraham2013}) and ages ($\sim$ 2-3 Myr, \citep{bik2012}. 
W3 Main is the most active star-forming region in the entire W3 complex, which 
contains more than 10 \hii  regions of various evolutionary status (\citealt{tieftrunk1997,ojha2004a,ojha2009}) and
 more than 15 massive stars of spectral type O3V-B0V \citep{bik2012} within the central 5.4 pc$^2$ area.
The central area of  W3 Main has cloud density as high as $\sim$ 2 $\times$ 10$^{23}$ cm$^2$, with two clumps of
masses 1700 and 800  M$_\odot$, respectively \citep{ingraham2013}. These values satisfy the conditions for high mass
star formation, which means that W3 Main may continue forming massive stars in the future.  Unlike W3 Main, at present AFGL333 
does not have the column density above the threshold for the formation of massive stars (i.e., N$_{H_2}$ $\sim$  1.8 $\times$ 10$^{23}$ cm$^2$; 
\citealt{krumholz2008,ingraham2013}). So far only one massive star of spectral type B0.5V has been identified within
AFGL333 \citep{hughes1982}.

For W3 Main, the SFE has been estimated as $\sim$ 44\% by \citet{bik2012} for an area $\sim$ 6.5 pc$^2$ around the core of W3 Main.
However, this is to be considered as an upper limit, as the cloud mass is estimated from $^{13}$CO and C$^{18}$O observations
(\citealt{dickel1980,thronson1986}),  where only the dense part of the region would have contributed to the cloud mass estimate.
Assuming the age of W3 Main  as 2-3 Myr \citep{bik2012}, the $\Sigma_{SFR}$ comes out to be
$\sim$ 110 - 160  M$_\odot$ Myr$^{-1}$ pc$^{-2}$, factor of $\sim$ 50 higher than that of AFGL333.

In spite of  being part of the same giant molecular cloud complex with similar size and age,
W3 Main and AFGL333 differ substantially in their level of star forming activities. 
%spezzi 2015
 This is similar to the  star formation rate variation observed among the most-studied nearby active star formation sites
in the Orion sub-regions, which includes regions such as ONC, L1630, L1641 etc. These clouds each 
have  masses (a few $\times$ 10$^4$ M$_\odot$) comparable to that of AFGL333. The star formation rate in L1630 is known 
to be a factor of 2 to 7 lower than that of  the nearby L1641,  despite having a very similar total reservoir of 
molecular material (\citealt{meyer2008,spezzi2015}).
One possible explanation  is the difference in the spatial distribution of dense gas between the two clouds  \citep{meyer2008}.
Similarly, W3 Main has a much larger reservoir of star forming gas in a compact area,  possibly formed by the
convergent constructive feedback scenario proposed by \citet{ingraham2013},  
which causes  the enhanced star formation rate when compared to AFGL333 (e.g., \citealt{gutermuth2011,lada2010}).
The cloud distribution as well as the star formation activities  within  W3 Main is similar to
that of the other high mass star forming regions such as NGC 6334 and W43,  whereas, AFGL333 resembles other  low mass star
forming regions. 

\subsection{Implications for triggered star formation in AFGL333}

Observational signposts of triggering in star forming regions include  pillar or cometary type structures protruding towards 
the massive stars, YSOs coinciding  with bubble rim/shells, age gradient between the YSOs located outside and inside the bright rim 
clouds, temperature gradient etc. (e.g., \citealt{chauhan2011,jose2013,pandey2013}).  The bright rim cloud BRC 5 is  located towards the southeast of AFGL333 and several 
short pillars extend into the W4  bubble interior. 
The winds and radiation emanating from young massive stars of W4 have  sculpted these elongated elephant trunks out of 
the surrounding molecular material. Based on an optical photometric analysis, \citet{panwar2014} reported that the YSOs located 
within the bright rim of BRC 5  are younger than  those YSOs outside the rim. Similarly, the dust temperature map from {\it Herschel} 
observations shows that the temperature is high ($\sim$ 22 K) near the ionization front  at BRC 5, and  is relatively low ($<$ 14 K)  
near the ridge \citep{ingraham2013}. The higher temperature at the ionization front indicates that it has been heated  by 
the strong radiation from W4. The above  examples are some of the classical signatures  of triggered star formation that are seen 
at the eastern edge of AFGL333. 

Triggered star formation can be defined in many ways, such as a temporary or long term increase in the star formation rate,
an increase in the final star formation efficiency or an increase in the total final number of stars formed \citep{dale2015}.
Our observational analysis shows that the star formation efficiency  and rate within AFGL333 is comparable to 
nearby low mass star forming regions. Stellar feedback has apparently not enhanced 
the star formation efficiency of  AFGL333. If  the massive OB stars of W4 had strong influence 
on  AFGL333, then  the closer region (AFGL333-main) should be more affected than the distant region (AFGL333-NW2).
However, we find almost similar ages,  star formation efficiencies  and star formation rates for the sub-clusters of AFGL333. 

Our analysis suggests that sequential  star formation, which  one would expect for a cloud  
under the influence of external ionizing photons, is not the prime mechanism of star formation in AFGL333.  
Star formation influenced by  external feedback does not seem to have propagated  throughout the cloud within AFGL333. Although individual YSOs at the tips of 
the small clouds (e.g. BRC 5, pillars)  might be triggered, their contribution to the overall star formation efficiency  and rate
 of the entire cloud is low. The star formation in these sub-clusters could have  started due to the  primordial cloud collapse. 
The ionization and shock front of the W4 bubble may have stalled at the east of AFGL333, giving the impression of 
triggering. 

We would like to remind the readers that the star formation parameters of AFGL333 are  the average properties of the three major stellar 
groups (see Section \ref{spatial}), which  comprise  sub-clusters of various evolutionary status, environments and external conditions.
Due to the limited sensitivity and spatial resolution of the data sets in this study, we do not segregate the different populations 
within AFGL333-main.  Hence we are unable to detect  any local triggering process acting in scales of $\sim$ 1.5 pc or less.  In conclusion, 
W4 \hii region appears to have little or no effect on the overall, averaged star formation  activity within AFGL333, except 
that  for the easternmost regions  that associated with BRC 5.

Disentangling triggered star formation  from spontaneous star formation  requires precise determination of  
proper motion and ages of individual sources \citep{dale2015}. Even if  we assume that  some triggering is going on at the eastern 
side of the AFGL333 cloud,   the overall effect of feedback from  W4  on the properties of the entire AFGL333 
cloud is low. Numerical simulations by \citet{dale2012} also show that stellar feedback may simultaneously enhance or suppress 
star formation and may  not have a strong effect on the overall star formation efficiency.
Detailed observational  studies of a large sample of  star forming regions are needed to strengthen the  hypothesis 
that feedback does not measurably affect the global star formation rate and efficiency.

\section{Summary}

The W3 giant molecular cloud complex is one of the most active massive star forming regions in the outer Galaxy. W3 Main, W3 (OH)
and AFGL333 are the major sub-regions within W3. This complex has been subject to numerous investigations as it is an excellent 
laboratory for studying the feedback effect from massive stars of the nearby W4 super bubble. The low mass stellar content of AFGL333
was  poorly explored until this study.

We analyzed  the deep $JHK_s$ observations  complimented with {\it Spitzer}-IRAC-MIPS observations   to unravel the
low mass stellar population  as well as to understand the cloud structure and star formation activity within AFGL333.
Based on the NIR and mid-IR colors, we identified  812 candidate  YSOs in this region, of which 99 are classified as Class I 
and 713 as Class II sources. The survey is complete down to $\sim$ 0.2 M$_{\odot}$. This  survey increases the census of 
 YSO members of the region by a factor $>$3 compared to  previous studies. The spatial distribution and  the stellar
density analysis shows that a majority of the candidate YSOs are located mainly within three stellar groups,  named as AFGL333-main,
AFGL333-NW1 and AFGL333-NW2. The disk fraction estimated within three stellar groups are $\sim$ 50--60\%.

Using  NIR data as well as CO based column desnity map, extinction maps  across AFGL333 is constructed in order  to 
understand the cloud structure as well as to estimate the cloud mass of the region.  
Combining the stellar mass with cloud mass, average star formation efficiency of the region has been estimated as $\sim$ 4.5\%
and star formation rate as $\sim$ 130 - 190 M$_\odot$  Myr$^{-1}$. The star formation rate density ($\Sigma_{SFR}$) measured within
AFGL333  is $\sim$ 2 - 3 $M_{\odot}$ Myr$^{-1}$pc$^{-2}$. 

We compared the star formation activity of AFGL333 with that of nearby low mass and high mass star forming regions as well as with W3 Main. 
The star formation rate density of AFGL333 is similar to the  nearby low mass star forming regions but is  a factor of $\sim$ 50 lower than
that of W3 Main. Currently AFGL333 is not dense enough to form massive stars.  On the other hand, the star formation 
activity within  W3 Main is comparable to other high mass star forming regions of the Milky Way and the region is still dense enough to form 
many more massive stars in future.  

Though we observe some of the classical signs of triggering such as pillars, bright rim cloud etc. towards the eastern edge of AFGL333,
we find no evidence to suggest that stellar feedback has influenced the global star formation activity within AFGL333.
 The star formation activity in AFGL333 and W3 Main are different most likely  due to the difference in the gas density within them 
as well as due to the differences in the feedback mechanisms in these two regions.
However, detailed studies of a large sample of externally influenced regions are essential in order to quantify this statement.  

\acknowledgments
We are grateful to the anonymous referee for his/her constructive comments that have helped us to improve the scientific contents of the paper.
The authors would like to  express our sincere gratitude to 
Robert Swaters for his help with NEWFIRM data reduction and Alana Rivera-Ingraham for sharing the column density map.
This work is supported by a Youth Qianren grant to GJH and general grant \# 11473005 awarded by the National Science Foundation of China.
This project is also based upon work supported by the National Science Foundation, USA, under Astronomy and Astrophysics Research Grant AST-0907980 to JSK.
The observations reported here were obtained at the Kitt Peak National Observatory, National Optical Astronomy Observatory, which is
operated by the Association of Universities for Research in Astronomy (AURA), Inc., under cooperative agreement with the National Science
Foundation.  This research has made use of the SIMBAD database (operated at CDS, Strasbourg, France),  Two Micron All Sky Survey 
(a joint project of the University of Massachusetts 
and the Infrared Processing and Analysis Center / California Institute of Technology, funded by NASA and NSF) and archival data 
obtained with the {\it Spitzer} Space Telescope (operated by the Jet Propulsion Laboratory, California Institute of Technology 
under a contract with NASA).

%\bibliographystyle{apj}
%\bibliography{ref}
%\bsp

\clearpage
\end{document}